\documentclass[fleqn,usenatbib]{mnras}

\usepackage{newtxtext,newtxmath}

\usepackage[T1]{fontenc}

\DeclareRobustCommand{\VAN}[3]{#2}
\let\VANthebibliography\thebibliography
\def\thebibliography{\DeclareRobustCommand{\VAN}[3]{##3}\VANthebibliography}

\usepackage{graphicx}	
\usepackage{amsmath}	
\usepackage{subfiles}
\usepackage{xcolor}

\usepackage{siunitx}
\DeclareSIUnit\year{yr}
\DeclareSIUnit\parsec{pc}
\DeclareSIUnit\msun{M_\odot}
\DeclareSIUnit\Rsun{R_\odot}

\newcommand{\kms}{\unit{\km\per\s}}
\newcommand{\kpc}{\unit{\kilo\parsec}}

\defcitealias{PP21.2021NatAs...5..251P}{PP21}
\defcitealias{Erkal_19_LMC_MASS.10.1093/mnras/stz1371}{E19}
\defcitealias{Erkal.modelrelease.2020}{E20}
\defcitealias{Lilleengen.oc.2023}{L23}
\defcitealias{Vasiliev.secondpassage.2024}{V24}
\defcitealias{GaravitoCamargo.model.2019}{GC19}
\defcitealias{2020MNRASReflexPP20}{PP20}
\defcitealias{Donaldson.2022}{D22}

\newcommand{\vtravel}{v_{\rm travel}}
\newcommand{\vecvtravel}{\vec{v}_{\rm travel}}
\newcommand{\lapex}{\ell_{\rm apex}} 
\newcommand{\bapex}{b_{\rm apex}}
\newcommand{\vr}{v_{r}}
\newcommand{\vphi}{v_{\phi}}
\newcommand{\vtheta}{v_{\theta}}

\newcommand{\ignore}[1]{}

\usepackage{pifont}
\newcommand{\cmark}{\ding{51}}%
\newcommand{\xmark}{\ding{55}}%

\title[radial variation of the MW disc reflex motion]{The radial variation of the LMC-induced reflex motion of the Milky Way disc observed in the stellar halo}

\author[Yaaqib, Petersen, Pe\~{n}arrubia]{
Rashid Yaaqib,$^{1}$\thanks{E-mail: rashid.yaaqib@ed.ac.uk}
Michael S. Petersen,$^{1}$
Jorge Pe\~{n}arrubia $^{1,2}$
\\
$^{1}$Institute for Astronomy, University of Edinburgh, Royal Observatory, Blackford Hill, Edinburgh EH9 3HJ, UK\\
$^{2}$Centre for Statistics, University of Edinburgh, School of Mathematics, Edinburgh EH9 3FD, UK}

\date{Accepted XXX. Received YYY; in original form ZZZ}

\pubyear{2024}

\begin{document}
\label{firstpage}
\pagerange{\pageref{firstpage}--\pageref{lastpage}}
\maketitle

\begin{abstract}
We measure the kinematic signature arising from the Milky Way (MW) disc moving with respect to the outer stellar halo, which is observed as a dipole signal in the kinematics of stellar halo tracers.
We quantify how the reflex motion varies as a function of Galactocentric distance, finding that (i) the amplitude of the dipole signal increases as a function of radius, and (ii) the direction moves across the sky. We compare the reflex motion signal against a compilation of published models that follow the MW--LMC interaction. These models show a similar trend of increasing amplitude of the reflex motion as a function of distance, but they do not reproduce the direction of the disc motion with respect to the stellar halo well. We also report mean motions for the stellar halo as a function of distance, finding radial compression in the outer halo and nonzero prograde rotation at all radii. The observed compression signal is also present in MW--LMC models, but the rotation is not, which suggests that the latter is not induced by the LMC. 
We extensively validate our technique to measure reflex motion against idealised tests. 
We discuss prospects for directly constraining the mass and orbital history of the LMC through the impact on the motion of the MW stellar disc, and how the modelling of the reflex motion can be improved as more and better data become available.
\end{abstract}



\section{Introduction}

The presence of the Large Magellanic Cloud (LMC) in the Milky Way (MW) halo complicates attempts to model the MW. Recent mass estimates of the LMC that include an extended LMC dark matter (DM) halo challenge equilibrium models of MW. Studies show that the LMC is in a first in-fall scenario\footnote{While other modeling of the MW--LMC system has found plausible second-passage scenarios \citep{Vasiliev.secondpassage.2024}, the results in this work remain unchanged, as we shall see below.} \citep[as found early on by ][]{Besla_kalliv_2020_first_infall} with a mass of $(1-3)\times10^{11} M_{\sun}$ \citep{Besla_Kalliv_2012,LMCTIMING_JP.2016MNRAS.456L..54P,Erkal_19_LMC_MASS.10.1093/mnras/stz1371}. Evidence for a heavy LMC calls for the development of non-equilibrium MW models, as observations are already sensitive to the time-dependence observed in recent models \citep{Rozier.etal.2022} and simulations \citep{GaravitoCamargo.model.2019,GaravitoCamargo.bfe.2021,Donaldson.2022,Lilleengen.oc.2023}. 

A recent example of a time-dependent MW halo model with observational impact is \citet{ORPHAN_KOPOSOV.2023MNRAS.521.4936K}, where the MW halo and disc were modelled as rigid bodies and allowed to move relative to the (rigid) LMC. This time-evolving potential was used to fit the Orphan (OC) stream. The result showed that the when allowing the MW to react to the infall of the LMC, the resulting time-dependent potential is able to match the OC stream velocity track on the sky (and a model without the LMC could not). Building on this work, a fully self-consistent modelling of the MW--LMC system was performed by \citet{Lilleengen.oc.2023}, where the time-evolving potentials of the MW and LMC were used to generate mock OC-like streams. In their OC-like streams, they found that the effects of time-dependent deformation in the MW and LMC DM halos on the stream observables were larger than observational errors. 
    
\citet{G_mez_2015} first showed that the presence of the LMC will cause the displacement of the MW disc from its equilibrium location at the barycentre of the total MW\footnote{For recent dark matter halo models of the MW, the barycentre is controlled by the dark matter, which outweighs the stellar disc by roughly a factor of 20.}. This displacement should be detectable as a dipole in velocity, when measured using halo stars with large dynamical timescales \citep[][PP20 hereafter]{2020MNRASReflexPP20}. \citet[][PP21 hereafter]{PP21.2021NatAs...5..251P} detected the LMC-induced displacement of the MW disc by measuring a signal in the observed galactic velocities of halo stars\footnote{Corroborated with line-of-sight velocity analysis in \citet{Erkal.2021.SLOSHING}.}, where the disc travel velocity was found to be $v_{\text{travel}}=32^{+4}_{-4}~\kms$, using a Galactocentric velocity dipole model, and using only the stars with large dynamical timescales by selecting stars at $r_{\text{galactocentric}}>40~\kpc$. In addition to the travel velocity, the direction of the disc motion (apex direction) was also recovered from the same model. The location of the apex was found to point along the previous trajectory of the LMC with $(\ell,\rm b)_{\text{apex}} = (56^{\circ}, -34^{\circ})$. Their result hints at the possibility that the there is valuable information of the trajectory of the LMC in the the velocity and the direction of disc motion. One of the implications of the work in \citetalias{PP21.2021NatAs...5..251P} is that studies of the MW stellar halo should take into account the reflex motion of the disc, as neglecting this will likely introduce biases in the kinematics of the halo stars \citep[see, e.g.][]{Erkal.modelrelease.2020,Deason.2021}. 

Following the work of \citetalias{PP21.2021NatAs...5..251P}, this paper investigates the dependence of the reflex motion signature on Galactocentric distance, motivated by the findings of \citetalias{2020MNRASReflexPP20}, who demonstrated that the LMC-induced dipole signal should vary with Galactocentric distance. We use the same velocity dipole model as in \citetalias{PP21.2021NatAs...5..251P} to characterise how the measured reflex motion varies with Galactocentric radius. At the radii used in \citetalias{PP21.2021NatAs...5..251P} ($r_{\text{galactocentric}}>40~\kpc$), the orbital timescale of stars is $\gtrsim$1 Gyr, which leaves little time for the outer halo stars to react to the in-falling LMC. In contrast, the inner halo will likely be responding to the LMC differentially with distance, including the limit where the innermost part of the halo will be following the disc potential, and exhibit no reflex motion signal. In the transition region between the inner most and outer halo, the reflex motion signature will be a complex function of the mass ratio of the LMC to MW, the orbital history, and the deformations of the MW and LMC.

In Section~\ref{sec:dataandmodels}, we describe the stellar halo data introduced in \citetalias{PP21.2021NatAs...5..251P}, and collect a range of models for the MW--LMC interaction. The comparison between models shows that different model assumptions produce different dipole signals. Therefore, the observational measurements can be used to further constrain MW--LMC models in future studies. Section~\ref{sec:methods} outlines the reflex motion model used and the fitting parameters, and describes the corresponding measurements made in the simulations. Our results are in Section~\ref{sec:results}. We present the measured the reflex motion at various distance bins between 20 kpc and 50+ kpc, and the direction of disc motion along with the modeled halo mean spherical velocities. The results show the variation of the directions and magnitude of disc motion. The results are compared to the various MW--LMC simulations described in Section~\ref{sec:dataandmodels}. Section~\ref{sec:discussion} discusses the observed radial dependence of the reflex motion parameters, and explores the possible dynamical justifications for the observed signal in the inner and outer halo. We describe the response of the disc when measured against stars at the large, intermediate and small distances. The limitations of the dipole model are assessed through various tests performed using mock catalogues generated from the simulations presented (see also the Appendices). Additional limitations of the sky coverage and tracer selection are discussed in this section. We conclude in Section~\ref{sec:conclusions}.
\section{Data and Models}
\label{sec:dataandmodels}

\subsection{Data}
\label{sec:data}

In this section, we briefly summarise the catalogue creation. We use the sample of K Giants and blue horizontal branch (BHB) stars assembled in \citetalias{PP21.2021NatAs...5..251P}. K Giants and BHBs, known to be good halo tracers, were identified using SDSS/SEGUE photometric and spectroscopic data\citep{SEGUE.Yanny_2009}. The positions and distances (and their uncertainties) are not modified beyond what is present in the SDSS/SEGUE data. A colour-cleaning technique was applied to the BHB sample to identify Blue Stragglers with the prescription of \citet{Lancaster.2019}. The BHB and K Giant catalogues were cross-matched with GAIA DR3\footnote{In \citetalias{PP21.2021NatAs...5..251P} Gaia DR2 was used, however, the catalogue was updated for this work.}, accepting stars with a reduced unit weighted error of {\tt ruwe}<1.4. The observed proper motions were converted to galactocentric coordinates with no loss of precision.

Sagittarius is the largest known substructure in the galactic halo and has members identified beyond galactocentric $r>20$ kpc. Members are present by eye when inspecting the parent sample. We perform the following cleaning of Sagittarius from the sample. We remove Sagittarius stream members from the sample by defining an angular momentum sphere of radius of 3000 kms$^{-1}$kpc around the location of the Sagittarius dwarf at $\vec{L}_{\rm sgr}$=(+605, -4515, -1267)km~s$^{-1}$~kpc \citep{SGR.GMM.PP21}. Stars in the plane of Sagittarius defined by \citet{Majewski.2003.2MASS.SGR} with $B_{\rm sgr} < 20^{\circ}$ that also fall in the sphere in angular momentum are tagged as Sagittarius members. Tests against the high-probability Sagittarius membership catalogue defined in \citet{SGR.GMM.PP21}\footnote{The catalogue is available at \url{https://github.com/michael-petersen/SgrL}.} suggest that no more than 10\% of the population remaining after the naive angular momentum cut are Sagittarius members.

In summary, the resulting dataset used in this work is identical to the individual stellar object samples of \citetalias{PP21.2021NatAs...5..251P} with the following differences: we use updated Gaia DR3 data and we include stars from $r>20~\kpc$.
The final dataset has the following number of stars: $N_{\rm K~giant} = 2830$, $N_{\rm BHB}=1445$. Finally, the data was binned in bins of 10 kpc between 20-50 kpc, halo stars with $r>50$ kpc were grouped in the same bin with a median distance in the bin of $\simeq59~\kpc$. Further binning beyond this radius leaves too few stars to fit. The final dataset consists of $N=4275$ (K Giants + BHBs), with $N = [2457, 1016, 446, 340]$ with bin centres of $r_{\rm galactocentric} = [25, 35, 45, 50+]$ kpc. In Appendix~\ref{appendix:b}, we test the effect of the SDSS sky coverage on our results. In Appendix~\ref{appendix:c}, we test the effect of fitting K Giants and BHBs separately. 
    
\subsection{Models}
\label{sec:models}

\begin{table*}
    \caption{\label{tab:models} Summary of MW--LMC models studied in this work.}
    \begin{tabular}{ccccccc}
        \hline
        Simulation & Self-Consistent? & $M_{\rm LMC,0}/M_{\rm MW}$ & MW halo profile & MW scale radius (kpc) & LMC halo profile & LMC $(\ell,b)_{r\to\infty}$ \\
        \hline
        \citetalias{GaravitoCamargo.model.2019}      & \cmark & 0.115 & Hernquist & 16.7$^{\rm a}$ & Hernquist & $\ell\simeq90^\circ,b>0^\circ$ \\ 
        \citetalias{Erkal_19_LMC_MASS.10.1093/mnras/stz1371} responsive      & \cmark & 0.158 & NFW & 18.5 & Hernquist & $\ell\simeq90^\circ,b<0^\circ$\\ 
        \citetalias{2020MNRASReflexPP20} 10\%    & \xmark & 0.1 & NFW & 14.1 & Plummer & $\ell\simeq90^\circ,b>0^\circ$ \\  
        \citetalias{2020MNRASReflexPP20} 20\%    & \xmark & 0.2 & NFW & 14.1 & Plummer & $\ell\simeq90^\circ,b>0^\circ$ \\ 
        \citetalias{Erkal.modelrelease.2020}      & \xmark & 0.15 & NFW$^{\rm b}$ & 16.0 & Hernquist & $\ell\simeq90^\circ,b\approx0^\circ$ \\ 
        \citetalias{Donaldson.2022} & \cmark & 0.238 & NFW & 15.0  & NFW & $\ell\simeq90^\circ,b<0^\circ$ \\ 
        \citetalias{Lilleengen.oc.2023} & \cmark & 0.158 & NFW & 12.8 & Hernquist & $\ell\simeq90^\circ,b<0^\circ$ \\ 
        \citetalias{Vasiliev.secondpassage.2024}    & \cmark & 0.273$^{\rm c}$ & NFW & 16.5 & NFW & $\ell\simeq0^\circ,b<0^\circ$$^{\rm d}$ \\ 
        \hline
        \multicolumn{7}{l}{The three different types of models are Hernquist \citep{Hernquist.1990}, NFW \citep{Navarro.Frenk.White.1997}, and Plummer \citep{Plummer.1911}.}\\
        \multicolumn{7}{l}{$^{\rm a}$The Hernquist scale radius is not the same measurement as the NFW scale radius. While \citetalias{GaravitoCamargo.model.2019} reports 40.82 kpc as the Hernquist scale length, the}\\
        \multicolumn{7}{l}{comparative measurement, where ${\rm d}\log \rho/{\rm d}r = -2$, is given here.}\\
        \multicolumn{7}{l}{$^{\rm b}$This MW halo is the same as that in {\tt galpy}'s {\tt MWPotential2014} \citep{Bovy.galpy.2015}.}\\
        \multicolumn{7}{l}{$^{\rm c}$At the time of the most recent pericentre, the mass is approximately half of the infall mass (cf. the ODE solution in Section 3.1 of \citetalias{Vasiliev.secondpassage.2024}).}\\
        \multicolumn{7}{l}{$^{\rm d}$As the \citetalias{Vasiliev.secondpassage.2024} are second-passage models, the location of the LMC as $r\to\infty$ is probably not the parameter controlling the location of the peak reflex signal,}\\
        \multicolumn{7}{l}{but we defer this investigation to future work.}
    \end{tabular}
\end{table*}

To connect the data to the dynamics of the MW--LMC interaction, we use a collection of published literature models. 
Broadly, there are two classes of simulations: idealised models that are highly efficient at probing dynamical principles, and self-consistent simulations that seek to capture the complete dynamics of the MW--LMC system. In Table~\ref{tab:models} we summarise the relevant parameters of the models.

\subsubsection{Idealised models}
\label{subsubsec:idealisedmodels}

\citetalias{2020MNRASReflexPP20} presented a suite of simulations featuring three different mass ratios (10\%, 20\%, and 30\% of the MW mass) of Plummer softened point mass LMCs introduced to a fiducial MW model. The LMC models were softened to match the  stellar rotation curve of the LMC constraint from \citet{vanderMarel.2014} while reaching the overall LMC target mass. These idealised models were run with {\sc exp} \citep{Petersen.exp.2022}, a flexible basis-function expansion code. Only the 10\% model was analysed in \citetalias{2020MNRASReflexPP20}, demonstrating that even at low mass ratios, the LMC is able to induce significant displacement of the MW barycentre -- resulting in observable reflex motion signals. Crucially, that paper also showed that the LMC-induced dipole signal varies with galactocentric distance. The models were later analysed in more detail in \citetalias{PP21.2021NatAs...5..251P}, demonstrating a trend in reflex motion strength with increasing LMC mass ratio (and validating the model technique; see also Appendix~\ref{sec:mockfits}).

To quantify the biases when measuring the overall MW mass from the LMC-induced distortion, \citet[][E20 hereafter]{Erkal.modelrelease.2020} analysed a stellar halo in a non-deforming MW DM halo potential as the LMC was brought in along a trajectory found by rewinding the LMC from present-day observables with a prescription to mimic dynamical friction.
This relatively simple model captures the bulk motion of the MW disc relative to the stellar halo. However, the signal is likely underestimated owing to the non-responsive nature of the MW DM halo.

The idealised models capture the response of the MW stellar halo, but should not be expected to either capture the true trajectory of the LMC or the deformation, dynamical friction, and mass loss of the LMC nor necessarily the response of the MW potential. However, these models are generally less computationally expensive to run, which enables larger numbers of particles and the ability to explore more models rapidly. An appropriate compromise for measuring reflex motion properties appear to be models that feature a deforming MW \citepalias[as in, e.g.,][]{2020MNRASReflexPP20}, even with a fixed LMC.

\subsubsection{Self-consistent models}
\label{subsubsec:selfconsistentmodels}

\citet[][GC19 hereafter]{GaravitoCamargo.model.2019} presented the first of the modern MW--LMC simulation suites. Their work explored the dynamical response of the MW DM halo to the passage of the LMC, finding that the response could be separated into two categories, the `local wake' (a manifestation of dynamical friction trailing the LMC) and the `collective wake' (a resonantly driven response that can span the entire DM halo). In addition to those deformation effects, \citetalias{GaravitoCamargo.model.2019} predicted a north/south reflex signature in line of sight velocities owing to the displacement of the MW disc. While the measurements below are in principle sensitive to the local and collective wakes, we expect the dominant signal to owe to the reflex signature --- an assumption we verify when calculating the all-sky reflex motion signatures resulting from the displacement of the MW disc with respect to the MW DM halo barycentre (Section~\ref{sec:results}). In this work, we analyse the `fiducial' model, their Model 7 \citep[see Table 1 in][]{GaravitoCamargo.bfe.2021}. This model features a radially biased MW halo and lightly rotating halo, which other works have suggested impacts the observed dynamical response \citep{Rozier.etal.2022}.

\citet[][D22 hereafter]{Donaldson.2022} presented a higher resolution version of the mass model used in \citet{PPJ.2022} for the purposes of investigating the DM distribution in the solar neighbourhood. These models were used to characterise the effect of deformation in both the MW and LMC at the solar neighborhood, and in particular compared local diagnostics when the MW or LMC was not responsive. In passing, \citetalias{Donaldson.2022} reported a reflex motion signal of 51 \kms, which is effectively a mean value for all halo particles (with the omniscience provided by the simulation).

\citet[][L23 hereafter]{Lilleengen.oc.2023} presented an updated self-consistent version of the best-fit spherical model proposed in \citet[][E19 hereafter]{Erkal_19_LMC_MASS.10.1093/mnras/stz1371}\footnote{See the `sph. rMW+LMC' parameters in their Table A1.}. This model features a modestly less concentrated MW halo than the originally fit \citetalias{Erkal_19_LMC_MASS.10.1093/mnras/stz1371} model, using $r_{\rm s}=18.5~ \kpc$ instead of $r_{\rm s}=12.8~\kpc$. In practice, this results in a modestly different LMC trajectory, but the response features are similar, including the reflex amplitude and direction of disc travel. For completeness, we have also produced a new version of the exact model parameters found for a reflexive spherical MW+LMC model in \citetalias{Erkal_19_LMC_MASS.10.1093/mnras/stz1371} (listed as `responsive' in Table~\ref{tab:models}). Both models were run with {\sc exp}, with simulation parameters describe in \citetalias{Lilleengen.oc.2023}. We can attribute the differences in the reflex amplitude to the concentration of the halo, but this appears to be a subdominant effect to the ratio of the LMC to MW mass. 
 
\citetalias{Lilleengen.oc.2023} quantified the deformation of the MW DM halo in their Figure~1, finding that the dipole is the strongest feature, but the quadrupole (which is not accounted for in our model) is also a non-negligible contributor. Future work can explore the extension of our model to encompass this signal.

\citet[][V24 hereafter]{Vasiliev.secondpassage.2024} ran a suite of models designed to probe the possibility of an LMC pericentre several Gyr before the most recent pericentre. As a representative model, we choose to compare to the model combining $\mathcal{L}3$ and $\mathcal{M}11$\footnote{We checked that the other models produce similar reflex motion signatures, so we do not include them in our analysis.}, resulting in an initial mass ratio of $M_{\rm LMC}/M_{\rm MW}=0.273$. However, the LMC model loses approximately half the infall mass on the earlier pericentric passage, which makes the most recent pericentre better approximated by a smaller mass ratio when comparing to single-passage models. 
To make a more fair comparison with first-infall models, we approximate the infall mass ratio as $M_{\rm LMC}/M_{\rm MW} = 0.138$, consistent with the value used for the ODE orbit approximation in Section~3.1 of \citetalias{Vasiliev.secondpassage.2024}.

Compared to idealised models, these self-consistent models are significantly more computationally intensive to run, a fact borne out by the reduced MW particle numbers in each run, making discreteness noise apparent in the curves of Figures~\ref{fig:apex} and \ref{fig:bulk_motion}, which we would expect to be smooth in the mean field limit. However, despite their lower resolution, the self-consistent models are crucial for resolving the full dynamical response of the MW and LMC owing to their interaction. One particular limitation of the current generation of MW--LMC models, both idealised and self-consistent, is a lack of cosmological context. In particular, all models we analyse in this work begin with spherically symmetric MW and LMC halos. The next generation of models should explore, at a minimum, the implications of triaxial halos for the reflex motion signal.
\section{Methods}
\label{sec:methods}

In this section, we describe the model parameters and fitting procedure used to measure the reflex motion from the data (Section~\ref{subsec:modeldescription}) and the corresponding measurements from the simulations (Section~\ref{subsec:measuringdipolesim}). 

\subsection{Model description and inference}
\label{subsec:modeldescription}

We employ the method used in \citetalias{PP21.2021NatAs...5..251P} here on the binned stars defined in Section~\ref{sec:data}. The analysis was performed using the six-dimensional phase space information of stars in our sample, where the data contains (for each star) galactic coordinates ($\ell,\rm b$), distance ($D$), proper motions ($\mu_{\ell}, \mu_{\rm b}$ and the correlation $\rho_{\mu_\ell\mu_{\rm b}}$) and line-of-sight velocities ($v_{\rm los}$).

As in \citetalias{PP21.2021NatAs...5..251P}, we fit an on-sky velocity model that contains nine free parameters. The first three parameters describe the reflex motion induced by the LMC: the magnitude of disc motion $\vtravel$ and apex directions $\lapex$, $\bapex$, which combine to make the three-dimensional vector describing the disc motion, $\vecvtravel$. The next three parameters accounts for any non-zero mean motion in the halo velocity through the bulk (mean) motion parameters $\langle \vr \rangle$, $\langle \vtheta \rangle$, and $\langle \vphi \rangle$. The final three model parameters are the hyperparamaters $\sigma_{\rm h, los},\sigma_{\rm h, \ell}$ and $\sigma_{\rm h, b}$.

In the Galactocentric coordinate system, we rotate the system such that the z-axis is aligned with the disc motion and points to $\lapex$, $\bapex$. To rotate the coordinate system, the following Euler rotation matrix was used to make the transformation\footnote{We use the $xyz$ convention for the rotation matrix.}
\begin{equation}
    \label{eq:Eul-rot}
    \left(\phi, \theta, \psi\right)_{\rm rot} = \left(\ell_{\rm apex}, \pi/2 - b_{\rm apex}, 0\right)
\end{equation}
where $\phi,\theta,\psi$ are the Euler rotation angles. In this frame the Cartesian representation of $\vecvtravel$ is simply $(0,0,-\vtravel)$ i.e. the reversed reflex motion of the MW disc w.r.t to the stellar halo.
Conversions between heliocentric Cartesian and Galactocentric Cartesian were performed by adopting a right-handed Cartesian coordinate system with the Sun positioned at $\vec{r}_{\odot \rightarrow MW}=(-8.3, 0.0, 0.02)~\kpc$ \citep{Gravity_collab_19,bennet_bovy_19}, with velocity $\vec{v}_{\odot \rightarrow MW}=(11.1, 244.24, 7.24)~\kms$ \citep{schonrich_binney_dehnen_10,Eilers_19_mwcircvel}. To find the average velocity vector of the halo using the reflex motion and bulk motion parameters defined above, the reflex motion model is represented by the sum of the dipole and bulk motion parameters,
\begin{equation}
    \label{eq:dipolemodel}
        \langle \vec{v}\rangle = \langle \vecvtravel \rangle + \langle \vr \rangle + \langle \vphi \rangle + \langle \vtheta \rangle - \vec{v}_{\odot \rightarrow MW}
\end{equation}
where $\langle\vec{v}\rangle$ is the mean velocity vector of the halo in the galactocentric frame. To transform $\langle\vec{v}\rangle$ to observed velocities, we project the velocities to galactic coordinates using the unit vectors aligned with line-of-sight and $(\ell,b)$ galactic coordinates \citepalias[equations 3 and 4 in ][]{PP21.2021NatAs...5..251P}.

In summary, the nine free parameters are the magnitude of disc motion $\vtravel$, the apex directions $\lapex, \bapex$, and the mean spherical velocities of stars $\langle \vr \rangle,\langle \vtheta \rangle, \langle \vphi \rangle$ and the hyperparameters for each of the observed galactic velocities $\sigma_{\rm h, los},\sigma_{\rm h, \ell}$ and $\sigma_{\rm h, b}$. In order to fit the model to the data, we obtain the posteriors on the model parameters by adopting a Gaussian likelihood. The likelihood is the product of two normal probability functions. The first is a 1D distribution for the line-of-sight velocity and the second is a 2D distribution for the proper motions containing also the covariance between the proper motions in the covariance matrix \citepalias[equations 5, 6 and 8 in ][]{PP21.2021NatAs...5..251P}. 
The hyperparameters are added in quadrature to the variances of the observed data and are used to minimise the covariance between the rest of the model parameters.

The fitting was performed using a nested sampling technique, which samples the posterior through computation of the evidence, via the Multinest \citep{Feroz_Hobson_Multinest.10.1111/j.1365-2966.2007.12353.x} package in Python \citep{pymultinest_buchner}. Multinest returns posterior samples and error estimates of the evidence. All priors are flat, with exception of the intrinsic scatter parameters which have inverse/Jeffreys priors. For a complete description of the likelihood used here, the reader is referred to the Methods section in \citetalias{PP21.2021NatAs...5..251P}.

\subsection{Measuring the Dipole Model Parameters from Simulations}
\label{subsec:measuringdipolesim}

We define here the procedure that was used to calculate the values of the reflex motion model parameters directly from the Cartesian 6D phase space information of the MW--LMC simulations. We do this in order to test the Bayesian inference code as well as to gain understanding of the physical meaning of the model parameters. Notice that since the mean velocity of halo particles is a combination of both the bulk motion of halo stars and the reflex motion of the MW disc, these parameters can be decoupled only by fitting them simultaneously. However, using only the 6D phase space information of stars from the simulation {\it without fitting}, it is not possible to fully decouple the bulk velocities ($\langle v_{\rm r,\phi, \theta}\rangle$) from the travel velocity as the reflex motion signal will project onto the measured velocity of halo stars.
Nonetheless, tests of fitting the simulation have shown that at all radii, the travel velocity dominates in $\langle \vec{v}\rangle$ in Equation~(\ref{eq:dipolemodel}). Therefore, in order to compute the apex directions and $\vtravel$, we simply compute the mean velocity of stars in each Cartesian direction using the halo stars within a given shell and transform to galactic coordinates. This is the definition of $\vtravel$ calculated {\it from the simulations}. 

For the bulk halo velocities, we follow a similar procedure, however, we must account for the reflex motion. For any halo star in the disc frame with $r_{\rm min} < r < r_{\rm max}$, where r$_{\rm min}$ and r$_{\rm max}$ are the inner and outer radii of any given shell, we compute the mean velocity in each Cartesian direction of stars in that shell (the reflex) and then subtract the value from the velocity of each individual star. In this definition of an inertial frame the new velocity is given by $v'_i = v_i - \langle v_i\rangle $ where $i=[x,y,z]$. Then we computed the spherical velocities of those stars and compute the mean spherical velocities of the halo in that shell. The bulk motions of the halo are defined in this frame that minimised aliasing of the reflex signal onto the spherical halo velocities.

In all the results of the reflex motion obtained from simulations, we note that the response of the halo is continuous with radius, and that calculating averages from shells may not be accurately representing the true underlying continuous distribution. With that said, the resolution of the simulation will almost always affect the choice of shell width, as more stars will lead to the ability to calculate the reflex and bulk motion parameters without running into discreteness noise. As we are presenting simulations with different particle numbers and techniques (and therefore resolution), we chose a bin width of 6 kpc, which is close to the bin width of 10 kpc in the data. We explore the effects of binning on the results in Appendix~\ref{appendix:d}.

\begin{figure}
    \includegraphics[width=\columnwidth]{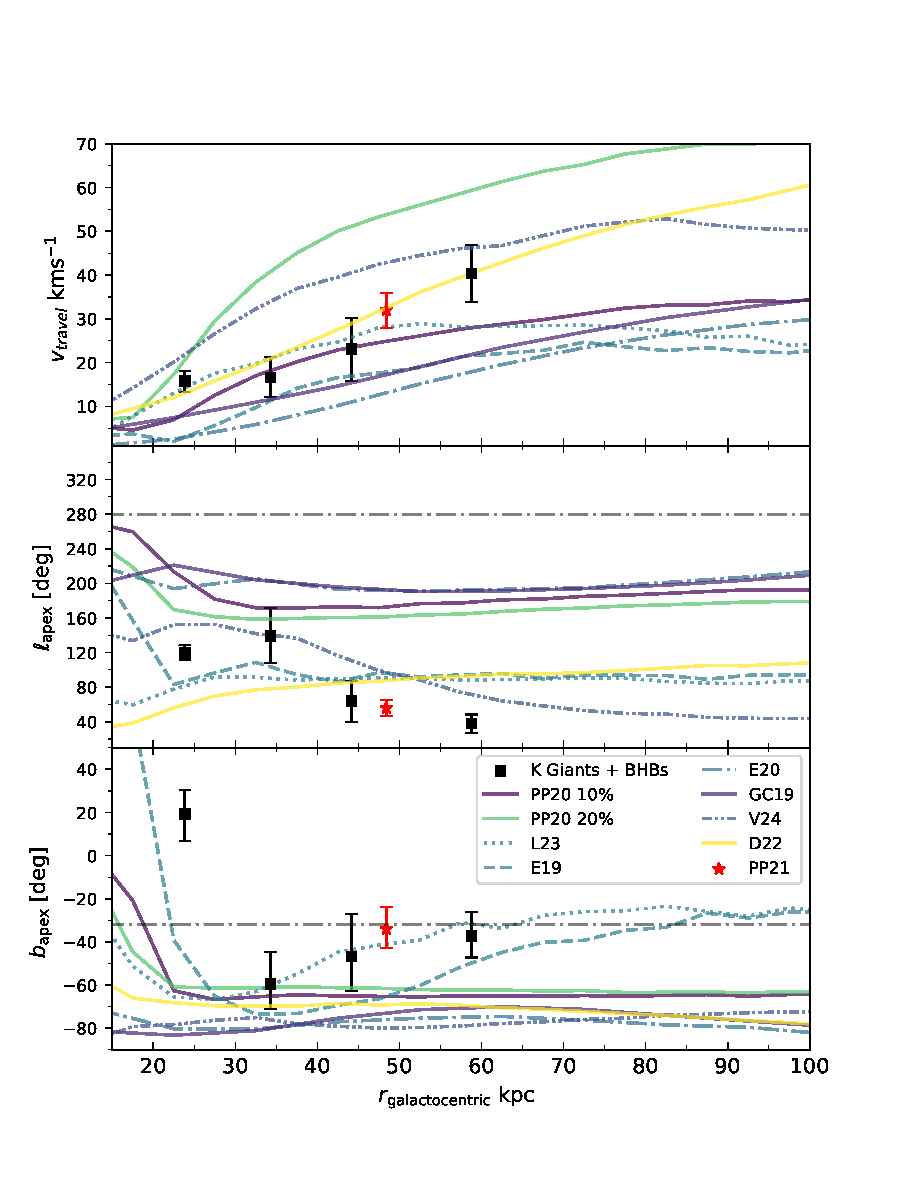}
    \caption{
    The measured values for the travel velocity and apex directions versus the median galactocentric distance of stars in each bin. Top panel: fitted travel velocity in each bin with uncertainties given as the standard deviation of the posterior chains of the parameters. Middle panel: Measured $\lapex$ values for stars in each bin, note that we restrict the apex longitude to be between 25$^{\circ}$ and 340$^{\circ}$. Bottom panel: Measured $\bapex$ values for stars in each bin, where the range of latitude angles are limited to $-90^{\circ}$ and $50^{\circ}$. In each panel, we plot corresponding measurements from the set of simulations (see Section~\ref{subsec:measuringdipolesim} for details in calculating the simulation curves). Colours of model curves correspond to the ratio of initial LMC mass to MW mass, with darker colours closer to 10\% and lighter colours closer to 25\%. The red point is the measured value for the parameters from \citetalias{PP21.2021NatAs...5..251P} using stars with r$>$40 kpc. The grey dash-dotted line shows the present day position of the LMC. Error bars indicate the $1\sigma$ width of the posterior distribution for each parameter. See Table~\ref{tab:models} for simulation names and mass ratios and Table~\ref{tab:result-summary} for the measured values and their uncertainties. The grey dot dashed lines mark the present day $(\ell,b)$ of the LMC.}
    \label{fig:apex}
\end{figure}

\begin{figure}
    \includegraphics[width=\columnwidth]{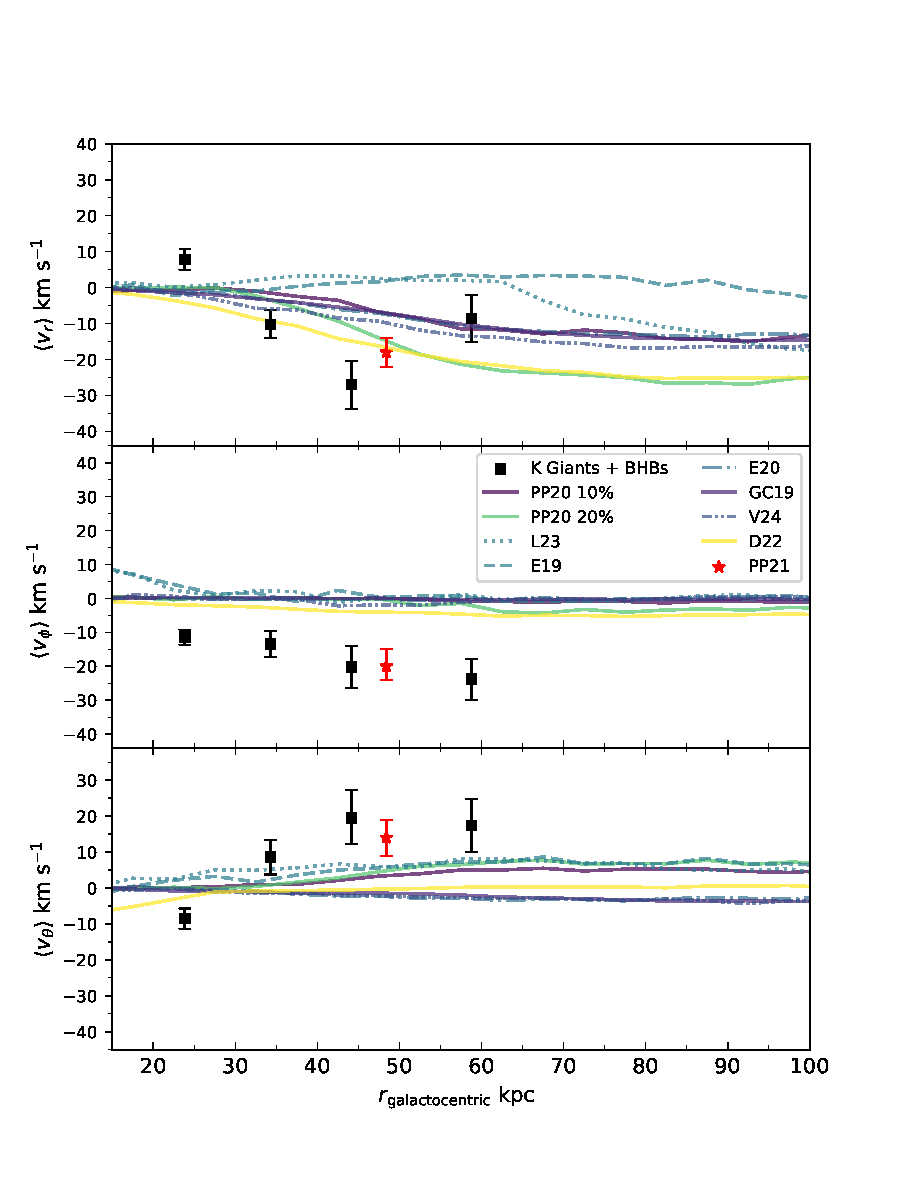}
    \caption{The measured halo bulk motion parameters as a function of galactocentric radius. Top panel: Mean halo motion in the radial direction. Middle panel: Mean halo motion in the azimuthal direction. Bottom Panel: mean halo motion in the polar direction. In each panel, we plot corresponding measurements from the set of simulations, which have been reflex corrected. The red point is the measured value for the parameters from \citetalias{PP21.2021NatAs...5..251P} using stars with r$>$40 kpc. Error bars indicate the $1\sigma$ width of the posterior distribution for each parameter. See Table~\ref{tab:models} for simulation names and mass ratios and Table~\ref{tab:result-summary} for the measured values and their uncertainties.}
    \label{fig:bulk_motion}
\end{figure}

\begin{figure}
    \includegraphics[width=\columnwidth]{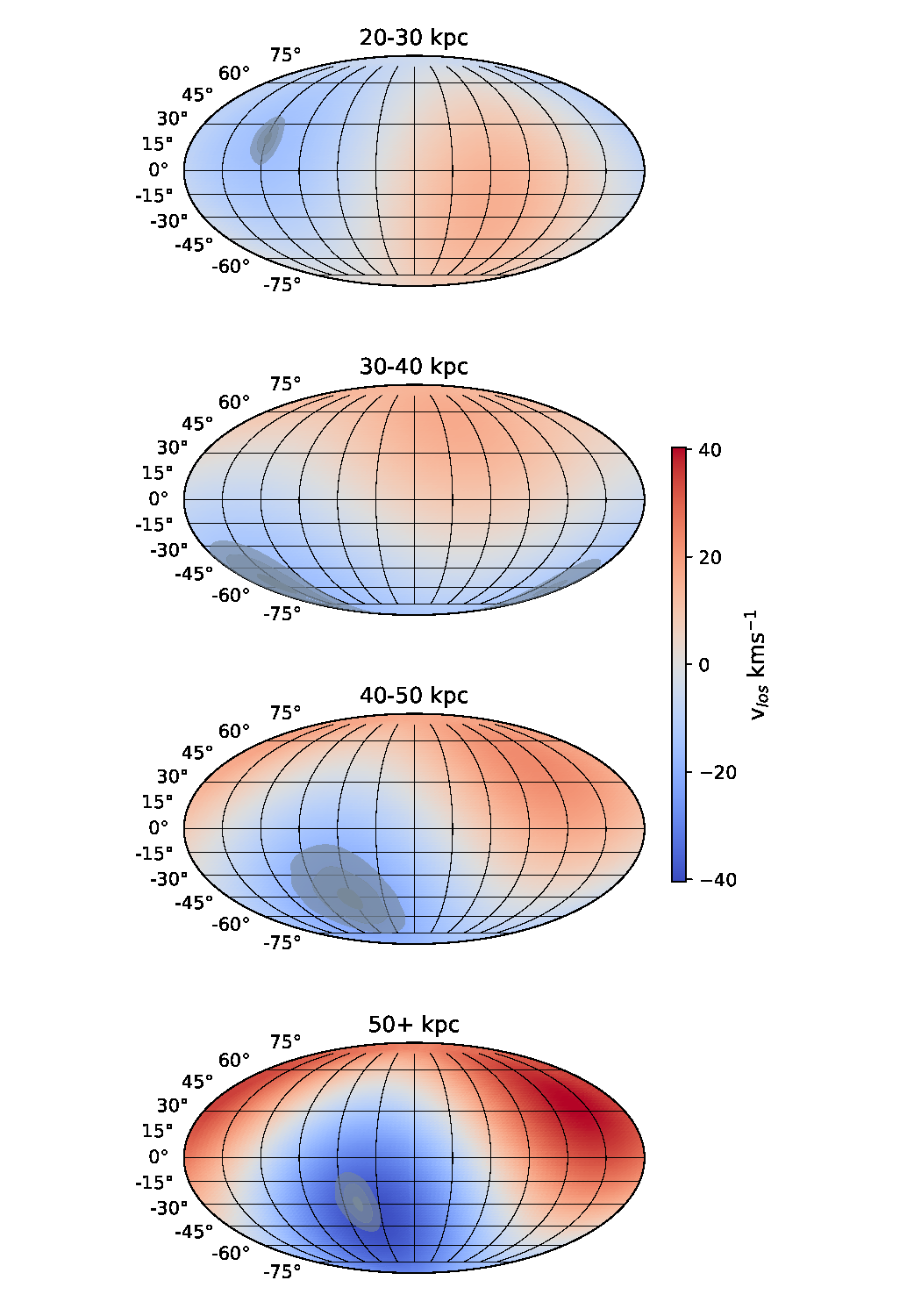}
    \caption{The on-sky line-of-sight velocity amplitude resulting from the dipole component of our reflex motion model, shown in Mollweide projection. Different panels correspond to different distances, from smallest radii (top) to largest radii (bottom). The amplitude of the signal increases with distance, and the apex location (with 67\%, 90\% and 95\% confidence ellipses shown in grey) moves across the sky (cf. Figure~\ref{fig:apex}) -- not tracing the orbit of the LMC, but rather the historical motion of the MW disc. }
    \label{fig:reflexmodel}
\end{figure}

\bgroup
\def\arraystretch{1.2}%
\begin{table*}
    \caption{\label{tab:result-summary} Summary of the dipole model fit results. All uncertainties on the values of the fitted parameters are the $1\sigma$ widths of the posterior distribution.}
        \begin{tabular}{l l lc c c c}
        \hline
        Description of Parameter      &parameters  &Units& 20-30 kpc & 30-40 kpc & 40-50 kpc & 50+ kpc \\ \hline
        Disc barycentric velocity     &$v_{\text{travel}}$  & \kms & 16$^{+2}_{-2}$ & 17$^{+5}_{-5}$ & 23$^{+7}_{-7}$ & 40$^{+7}_{-7}$   \\ 
        Apex longitude                &$\ell_{\text{apex}}$ & deg& 120$^{+9}_{-8}$ & 139$^{+32}_{-31}$ & 64$^{+23}_{-24}$ & 38$^{+11}_{-11}$   \\
        Apex latitude                &$b_{\text{apex}}$    & deg & 19$^{+11}_{-12}$ & -59$^{+15}_{-12}$ & -47$^{+20}_{-16}$ & -37$^{+11}_{-10}$  \\ 
        Mean halo radial velocity     &$\langle v_{r} \rangle$       & \kms & 8$^{+3}_{-3}$ & -10$^{+4}_{-4}$ & -27$^{+7}_{-7}$ & -9$^{+7}_{-6}$  \\ 
        Mean halo azimuthal velocity  &$\langle v_{\phi} \rangle$    & \kms & -11$^{+2}_{-2}$ & -13$^{+4}_{-4}$ & -20$^{+6}_{-6}$ & -24$^{+6}_{-6}$  \\ 
        Mean halo polar velocity      & $\langle v_{\theta} \rangle$ & \kms & -9$^{+3}_{-3}$ & 9$^{+5}_{-5}$ & 20$^{+8}_{-7}$ & 17$^{+7}_{-7}$   \\ 
        Line-of-sight velocity hyperparameter      &$\sigma_{\rm h,los}$ & \kms& 103$^{+2}_{-2}$ & 98$^{+2}_{-2}$ & 100$^{+4}_{-4}$ & 87$^{+4}_{-4}$  \\ 
        Galactic longitude velocity hyperparameter &$\sigma_{\rm h,\ell}$     & \kms& 76$^{+1}_{-1}$ & 78$^{+2}_{-2}$ & 85$^{+4}_{-4}$ & 70$^{+5}_{-4}$  \\ 
        Galactic latitude velocity hyperparameter  &$\sigma_{\rm h,b}$     & \kms& 61$^{+1}_{-1}$ & 83$^{+3}_{-2}$ & 96$^{+4}_{-4}$ & 74$^{+5}_{-5}$  \\ \hline
        Median distance of stars & & kpc & 23.85 & 34.31 & 44.14 & 58.80 \\
        Number of stars & & & 2461 & 1026 & 446 & 340  \\
 \hline
    \end{tabular}
\end{table*}
\egroup

\section{Results}
\label{sec:results}

The \citetalias{PP21.2021NatAs...5..251P} result showed that the MW disc was not at rest with respect to the outer halo. This first measurement of reflex motion found that the disc travel velocity was $\sim30~\kms$ in a sample of halo tracers with distances of $r>40~\kpc$\footnote{The stellar measurements were made using the same sample of K giants and BHBs as in this work; the satellite galaxy sample used in \citetalias{PP21.2021NatAs...5..251P} is not large enough to be divided into bins, as we do here. We therefore do not use this sample. Similarly, the MW globular clusters are largely located in the inner MW, where the reflex motion signal is expected to be small.}. Building on this work, we find that a single measurement of the reflex motion does not provide sufficient constraining power when comparing to self consistent MW--LMC models -- see the red data points in Figures~\ref{fig:apex}, which are the measurements from \citetalias{PP21.2021NatAs...5..251P}. We find the parameters of the dipole model described in Section~\ref{sec:methods} change when measuring the signal using stars at varying galactocentric radii. Furthermore, reflex motion measured directly from idealised and self-consistent MW--LMC simulations (see Section \ref{sec:models}) are shown alongside the data results to contrast existing models against observations. 

In this Section, we report the measurements of the reflex motion dipole in Section~\ref{subsec:vtravel-result}, the measurements of bulk velocities in Section~\ref{subsec:bulkmotion-result}, and compare the results with simulations in Section~\ref{subsec:simulationcompare-result}.
    
\subsection{Radial Variation of the Travel Velocity}
    \label{subsec:vtravel-result}

The likelihood defined in Section~\ref{sec:methods} is a powerful way to decouple the kinematic response of the MW disc and internal motion of the halo stars, without imposing a physical model for the MW--LMC system. The radial variation of $\vecvtravel$ shows {\it in what direction} and {\it how fast} the disc is travelling, relative to an ensemble of halo stars -- here, shells of halo stars. We present here the measured magnitude and direction of disc motion. Comparison against theoretical models suggests that response of the MW disc with radius depends on the MW and LMC mass profiles. Importantly, the direction of disc motion will also depend strongly on the infall trajectory of the LMC. 

Our likelihood analysis indicates that the reflex motion amplitude, $\vtravel$, generally increases with galactocentric radius in the data. The top panel of Figure~\ref{fig:apex} shows the resulting measurements from the stellar halo sample as black points. We also find that the measured on-sky direction of disc motion changes with radius. The lower two panels of Figure~\ref{fig:apex} show the on-sky location of the apex in Galactic coordinates $(\ell,b)$.
We find that $\lapex$ is in the opposite half of the sky from the present-day location of the LMC at 20-30 kpc then {\it becomes closer} in angular separation to the present-day LMC location in the more distant bins\footnote{Note that in Figure \ref{fig:apex}, the points seem be further away in distance, however considering the wrapping of the angular coordinate, the points approach the present day position of the LMC in terms of on-sky separation -- highlighting the need for fits at different radii.}. The apex longitude, $\lapex$, continues to slowly decrease, scanning across the sky, to a minimum value of $\sim 40^{\circ}$ in the final bin containing only stars with r$>$ 50 kpc. As expected from the number of stars in each bin, the uncertainties on the fitted $\lapex$ parameters are smallest in the 20-30 kpc bin with a $1\sigma$ uncertainty of about 10$^{\circ}$. In the following bin at 30-40 kpc, we find that the errors are roughly 4 times larger, much larger than expected when considering the distance errors and the number of stars are not too small in that bin, complicating robust inference of how much the $\lapex$ location may really be changing.
    
The apex latitude $\bapex$ is, surprisingly, measured to be in the Northern galactic hemisphere in the closest bin, then points toward the Galactic South in the subsequent bins. However, the location is always measured to be $\bapex>-60^\circ$, suggesting that rigid models treating the reflex motion as purely downward at a fixed direction are too simplistic to capture the observed behaviour. The uncertainty in the fitted $\bapex$ values does not vary greatly between the bins with $1\sigma$ widths of approximately $10^{\circ}$. The small change in the uncertainties on the fitted values is likely due to the dependence of the apex latitude fit on the sky coverage of the sample\footnote{In Appendix~\ref{appendix:b}, we explicitly test the effect of sky coverage.}.    

In summary, all three $\vecvtravel$ parameters change with radius: $\vtravel$ increases gradually with radius, and the apex angles approach the present-day location of the LMC with increasing distance, encoding information about the trajectory of the LMC.

\subsection{Radial Variation of the Bulk Motion parameters}
\label{subsec:bulkmotion-result}

In addition to the dipole model, we measure the `bulk motion' of the stellar halo, fit as three spherical velocity components with an all-sky mean value: the mean radial velocity $\langle v_r\rangle$, the mean azimuthal velocity $\langle v_\phi\rangle$ (cylindrical rotation), and the mean polar velocity $\langle v_\theta\rangle$. The mean velocities describe the velocity signal of the halo {\it distinct from} the reflex dipole. In addition to describing the response to the LMC infall, the mean velocities also describe any pre-existing ordered motion within the shells after removing the reflex dipole.
Perturbations such as a heavy LMC infall will affect not only the disc, but also the internal kinematics of the halo stars. The mean velocities in the halo could also encode any intrinsic rotation that the halo could have had prior to the infall of the LMC. The crucial step in our work when measuring the bulk motions through the likelihood approach is that the reflex motion of the disc is robustly accounted for when measuring the mean motion parameters. 

The mean radial velocity $\langle v_r\rangle$ (see Figure~\ref{fig:bulk_motion}) of the bulk halo motion show statistically-significant non-zero amplitude in the data, which varies with radius. 
In the innermost bin, we measure $\langle \vr\rangle =+8~\kms$ with the positive sign meaning that the halo is expanding. Moving outward, $\vr$ decreases to a minimum value (but maximum amplitude) of $\vr=-27~\kms$ in the distance bin of 40-50 \kpc, i.e. the halo is contracting at present. In the farthest bin, the result is consistent with no radial motion within the $1\sigma$ posterior width of the fitted value. The uncertainties on the fitted bulk motions increase by about 1km$~\rm s^{-1}$ at each successive bin, and they are the largest in the last bin, as would be expected for a decreasing number of stars in each successive bin.

For the mean azimuthal velocity $\langle v_\phi\rangle$, we find a consistent rotation trend with increasing galactocentric radius between 20-50+ kpc, with maximum amplitude of $\vphi=-24^{+6}_{-6}~\kms$ at the 50+ kpc bin. This rotation signal was measured to be statistically significant and prograde (in the same sense as the disc rotation) at all radii. Finally, the bulk polar motion in the shells of halo stars is largely consistent with zero velocity.

Using the bulk motion parameters of the dipole model, the results from the data show that the MW halo is experiencing compression, whose effect is greatest on halo stars between 40-50 kpc. The data also show evidence of mild cylindrical rotation with a prograde direction at all radii.

\subsection{Dipole Model Parameters in Simulations}
\label{subsec:simulationcompare-result}

The eight MW--LMC simulations collected for this work largely show the same results: a strong dipole signal, with a well-defined on-sky apex, and minimal bulk (mean) velocity signals caused by the LMC infall. Further, the models exhibit similar trends with radius. At 100 kpc, the models have largely reached the `impulsive' regime where the dipole signal becomes distance invariant \citepalias[cf.][]{2020MNRASReflexPP20}. The similarity of the model predictions poses a challenge for distinguishing between models using current-generation reflex motion measurements, this will be studied in a follow-up contribution. 

For $\vtravel$, in the intermediate bins (30-50kpc) the \citetalias{2020MNRASReflexPP20}, \citetalias{GaravitoCamargo.model.2019}, \citetalias{Erkal_19_LMC_MASS.10.1093/mnras/stz1371}, \citetalias{Donaldson.2022} and \citetalias{Lilleengen.oc.2023} models are consistent with the magnitude of $\vtravel$ within $1\sigma$, but do not show the same agreement in the 20-30 and final 50+ kpc bin.
All live MW--LMC simulations show a flattening of $\vtravel$, however, the radius at which the profile begins to flatten is not consistent between the simulations. Furthermore, the slopes of $\vtravel(r)$ vary between the simulations with \citetalias{Erkal.modelrelease.2020} showing an almost linear increase between 20-100 kpc while others show a profile that increases more rapidly and then flattens out at large radii. This difference in profile likely corresponds to the model choices (e.g. profile, mass, etc.) made for the LMC and MW in those simulations.
Additionally, the \citetalias{2020MNRASReflexPP20} models for the LMC show a clear difference in $\vtravel$ with mass, where the assumed MW--LMC mass ratio increases the magnitude $\vtravel$ at any fixed radius, but also sets where the profile flattens. The difference, in the form of an apparently exaggerated signal, in the idealised models from \citetalias{2020MNRASReflexPP20} relative to the self-consistent models suggests that the response of the LMC to the Galactic tidal field may be crucial to appropriately modelling the reflex motion.

For $\lapex$, there is a large variation between the simulations. In the inner halo at r$<$40 kpc, the behavior of the live and simple \citetalias{2020MNRASReflexPP20} simulations is not consistent, suggesting that further detailed modeling is required. 

In terms of the on-sky apex location relative to present-day location of the LMC, the \citetalias{Donaldson.2022}, \citetalias{Vasiliev.secondpassage.2024}, \citetalias{Lilleengen.oc.2023} and \citetalias{Erkal_19_LMC_MASS.10.1093/mnras/stz1371} are approximately 180$^{\circ}$ away from the LMC, while the \citetalias{Erkal.modelrelease.2020} and \citetalias{2020MNRASReflexPP20} models are somewhat closer to the present-day location of the LMC at approximately 90$^{\circ}$ away. All first-passage models do not show significant variation with radius at distances beyond 30 kpc, but all have an abrupt change in the $\lapex$ profile at small distances. We note two apparent families of behaviour of $\lapex$ in the models, where the longitude points at two different quadrants of the sky in the simulations. If the LMC is indeed on a second passage, as modeled in \citetalias{Vasiliev.secondpassage.2024}, the earlier infall of the LMC may `pre-process' the halo in ways that our dipole model does not fully capture. 

For the apex latitude, we find that \citetalias{GaravitoCamargo.model.2019}, \citetalias{Erkal.modelrelease.2020} and \citetalias{Vasiliev.secondpassage.2024} show almost no variation in $\bapex$ as a function of radius -- pointing largely at the Galactic South pole. Given the present-day latitude of the LMC of $b=-32^{\circ}$, almost all models do not point at that location, but are offset in $\bapex$ consistently where $\bapex$ remains constant at large (r$>40$kpc) distances. However, \citetalias{Erkal_19_LMC_MASS.10.1093/mnras/stz1371} and \citetalias{Lilleengen.oc.2023} both show similar profiles in $\bapex$, where in the inner halo the apex latitude becomes more negative (or decreases from positive to negative), then increases slowly with increasing radius, up to about 90 kpc where the profile begins to flatten. 
    
None of the simulations are in agreement with the data results in all six parameters simultaneously.
There are some live models models that agree well with the data in $\vecvtravel$ such as the \citetalias{Donaldson.2022} and \citetalias{Lilleengen.oc.2023}.
The largest discrepancy is seen in the bulk motion parameters, where the data show strong contraction in the halo between 20 and 50 kpc. Furthermore, the data show that there is a nonzero rotation signal in the bulk motion of the halo shells at all radii. Of the collected simulations, only \citetalias{GaravitoCamargo.model.2019} includes any ordered motion in the stellar halo pre-LMC infall (in the form of modest nonzero halo spin), but does not match the observed values. Whether this mismatch of bulk motion parameters can tell us something about the ordered motion in the pre-LMC infall MW halo is a subject for future study. Another issue we identify is that the `downward' models \citepalias{GaravitoCamargo.model.2019, Erkal.modelrelease.2020, Vasiliev.secondpassage.2024} -- where the LMC-induced reflex is approximated by a constant motion in the $-z$ direction relative to the MW disc -- are apparently inconsistent with the apex locations we find in the data. As the apex points to the part of the halo that appears to be moving away from us most quickly at a given radius, our findings support the idea that a disc motion does not point close to the galactic south pole. The live simulations presented here are able account for the magnitude of disc motion $\vtravel$ well at most radii.

\section{Discussion}
\label{sec:discussion}

As shown in Section \ref{subsec:measuringdipolesim}, ignoring the reflex motion dipole response induced by the LMC does not properly describe the net velocities in the stellar halo and may introduce bias in the measured bulk stellar halo velocities. This bias can affect measured properties of the MW halo \citepalias{Erkal.modelrelease.2020}. The radial variation of the travel velocity of the disc provides an avenue to constrain the mass ratio of the MW--LMC interaction. In the scenario of the disc barycentre dislodging from the (previously) shared disc + halo barycentre, the radial reflex motion signature will depend on the profiles of the MW and LMC, as well as on the trajectory of the LMC\footnote{The trajectory itself also depends primarily on the mass profile of the MW.}. The measurements made in this work put forward a way to test whether models are able to capture (in the same parameter space) the effect of the LMC's infall on the MW.

Recent studies have shown that the LMC on first infall may affect the trajectories of stellar streams \citep[e.g.][]{Shipp.etal.2019,ORPHAN_KOPOSOV.2023MNRAS.521.4936K}, and thus bias the constraints on the Galactic potential derived from models that adopt a static MW halo \citep[e.g.][]{Arora.etal.2023,Brooks.etal.2024}. In this context, it is worth noting that measuring the DM profile of the MW using the Sagittarius stream has previously yielded contradictory results, e.g. \citet{Law_2010.halo_sgr_oblate} find an oblate halo potential, whereas \citet{Fardal_2005,Johnston_2005} find a prolate one. In contrast, \citet{Bovy_2016} find a spherical shape by modelling GD-1. While \citet{ORPHAN_KOPOSOV.2023MNRAS.521.4936K} have improved the fitting method by allowing the MW disc and halo to move as rigid bodies when modelling the effect of the LMC's infall on the OC stream, \citetalias{Erkal_19_LMC_MASS.10.1093/mnras/stz1371} found that the OC stream's track can still be fit using either of the three aforementioned shapes with equal accuracy.

As the LMC loses mass in the tidal field of the MW during infall, its present-day bound mass is the largest factor setting the magnitude of the observed reflex motion \cite[compare, e.g., the first infall models with][where the LMC loses half of its mass in the earlier pericentric passage]{Vasiliev.secondpassage.2024}. Measuring the stellar halo properties of the MW will need to correct for reflex. We show in Section~\ref{sec:results} that the magnitude of disc velocity relative to the halo is non-zero even in the 20-30 kpc shell. 
Based on our analysis of a suite of different MW--LMC simulations, we propose that using simulations alongside dipole model measurements allowed to vary with radius may be a robust method to test assumptions about the {\it pre-infall} profile of the MW and LMC. Note that as the method shown in Section \ref{sec:methods} describes the mean velocity of the MW stellar halo as the sum of reflex and bulk motion, a model-data comparison must fit the stellar halo as a whole, rather than only modelling the effects on individual streams. 

\subsection{The Inertial Frame and the Adiabatic and Impulsive Regimes}

The reflex motion signature ($\vecvtravel$) is best probed using halo tracers with large apocentres and long dynamical times \citepalias{2020MNRASReflexPP20}, which have not yet responded to the LMC's infall. However, halo tracers at those distances will also be affected by the apparent reflex motion of the disc when observed from the disc frame. The likelihood approach used in \citetalias{PP21.2021NatAs...5..251P} and here decouples the velocity induced by the motion of the disc and the intrinsic motion of the halo through the modelling of the mean halo velocity as a sum of reflex and bulk motion.
Measuring the reflex motion at smaller distances (where the orbital times of stars is $\lesssim$1~Gyr, at distances $\lesssim$40~\kpc) shows that the magnitude of the travel velocity is smaller than at the most distant bin owing to the ability of the inner halo to respond differentially with radius to the gravitational pull of the LMC. 

Deep in the MW potential well, the orbital times are sufficiently short compared to the LMC perturbation such that there is no differential effect with radius. This is because the perturbed orbits have time to mix in phase-space, such that any differential effect with radius is washed out. This is the {\it adiabatic regime}, where the magnitude of disc motion is minimal. In this regime, the inner Galactic potential is dominated by the disc, and the deep potential at this region means that there is effectively no LMC deformation. If the assumption of the inertial Galactocentric frame is made at an earlier time (before the LMC’s infall), then reflex motion measured today using the most distant halo stars reflects the motion of the disc centre w.r.t. a pseudoinertial frame\footnote{`Pseudoinertial' here means accepting the frame of a shell of halo stars at very large distances (e.g. 100 $< r< 200$ kpc) as nearly matching the pre-infall inertial frame. We will refer to this below as the `inertial' frame.}. At very large distances, stellar orbits have large orbital times and their dynamical response to the LMC has not fully manifested. The barycentre of these stars is likely still that of the pre-infall MW disc + halo system. Therefore, outside of some large radius, the reflex motion signal may be approximated as an instantaneous kick. Using the outermost stars in the halo (at r$>50~\kpc$) the motion of the disc appears to be impulsive (the {\it impulsive regime}), and the travel velocity measured using stars in the inertial frame will be maximal. The transition from an adiabatic regime to an impulsive regime indeed depends on the mass profiles of the MW and LMC, and this dependence arises from how the differential response of the MW halo varies with radius. This is why the reflex motion measurements presented in this work provide an avenue for testing the differential response of the halo and how it depends on the orbital period of individual halo stars.

The various idealised MW--LMC simulations shown in Figures~\ref{fig:apex} and \ref{fig:bulk_motion} (see also Table~\ref{tab:models} for details), show that the current observational constraints are not distant enough to have reached an impulsive regime. Apex angles and mean spherical velocities in particular do not show evidence for radial convergence. Furthermore, we find that the reflex motion parameters measured directly from the simulations do not agree with the data. Each model shown in this work uses different assumptions about the profile of the LMC and the MW, although the majority adopts an NFW profile for the DM halo of the MW \citepalias[][being the exception]{GaravitoCamargo.model.2019}. 
We see the effect of the various assumptions leading to different predictions for $\vtravel$ and the apex angles. Interestingly, while models like 
\citetalias{Erkal_19_LMC_MASS.10.1093/mnras/stz1371} use the OC stream to constrain the profiles of the LMC and  MW, the resulting reflex motion parameters do not reproduce the reflex motion of the MW disc. These results call for a future theoretical efforts to fit stream motion-track and the reflex motion of the disc simultaneously.

\subsection{Halo Compression and Rotation}

Our analysis of the motion of halo stars gives evidence for the first time that the halo may be contracting over a wide distance range. This is likely induced by the dynamical response of the Galaxy to the LMC's infall.This effect is relatively straightforward to understand.
The contribution of stripped mass from the LMC deepens the potential of the MW, and the response of MW halo tracers is a net inward motion as a result of the added mass. Both simulations and data show this effect. This is particularly visible in simulations that model the LMC as a static Plummer potential in \citetalias{2020MNRASReflexPP20}, where there is no mass loss from the satellite (see Appendix~\ref{sec:mockfits}). We see a clear compression signal with $\langle\vr\rangle<0$. The amount of mass bound to the LMC at the present-day correlates with the amount of net inward motion that halo stars experience, and matching $\langle\vr\rangle$ may be crucial for constraining the present-day LMC mass. 

A surprising result from our data analysis is a clear evidence for rotation in the bulk motion of halo stars, which is non-zero at most radii at the $3-4\sigma$ level. None of the live models show a significant rotation. Future models may need to include prescriptions for intrinsic rotation in the halo to test whether this can produce the observed signal. 

\subsection{Limitations of the Dipole Model}

The dipole model used here to model the mean velocity of the halo as a sum of reflex and bulk motions at various radii is just the first-order expansion beyond the monopole, which studies have predicted or found higher-order signals \citep{cunningham.2020.responseharmonic,conroy.2021.response,Lilleengen.oc.2023}. The dipole model could be straightforwardly be extended to quadrupole deformations such as those proposed by \citet{GaravitoCamargo.model.2019}, and observed by \citet{conroy.2021.response}. The quadrupole expansion can be defined in a similar manner to the dipole model used here. Higher-order expansions will be necessary for future data releases that will increase the total number of stars in the sample. Additionally, this type of modelling of the velocity field may be a model-agnostic way of approaching substructure in the data \citep{cunningham.2020.responseharmonic}.

Future modelling efforts may want to avoid binning the data. This may require a prescription for the radial dependence of the parameters in the likelihood. Including this would add more fitting parameters, but it would be worth exploring as the number of distant halo stars with known 6D phase-space information grows in size and the astrometric uncertainties improve. Given the limitations of current data sets, we opted for binning the stars in radial shells. Some non-parametric alternatives may include the use of splines or other smoothing regressions. A more complex mean halo velocity model (e.g. equation \ref{eq:dipolemodel}) that includes a quadrupole term and radial term will increase the number of parameters significantly, which would require more data than what was used in our work. However, we also show that the trends observed in the data are not significantly affected by the choice of bins (see Appendix~\ref{appendix:d}). The affect of the binning procedure is akin to averaging the assumed continuous distribution in radius of the parameters.

With the current generation of MW--LMC models we are starting to explore the effect of the mass profiles on the reflex motion signature. Future work could probe the amount and nature of the DM particles \citepalias{GaravitoCamargo.model.2019}, as well as modelling the mass profile and shape of the MW and LMC simultaneously in order to match the observed radial variation of the reflex motion parameters. 
Here we also use sub-sampling of the DM halo to make tracers of the Galactic potential. Given that the stellar halo is thought to be made of debris from accreted satellites, hierarchical models are needed to determine whether or not a clumpy distribution of halo stars in the integral-of-motion space could affect the reflex motion measured in this work.

Finally, there is without doubt a need to correct for the motion of the disc when studying the kinematic structure of the MW halo. The apparent velocity induced by the motion of the MW disc will affect observations of the velocities of halo stars, most significantly beyond 40~\kpc. In particular, more nuanced measurements of the stellar halo ordered motions and anisotropy will be assisted by reflex corrections \citep[e.g.][]{Deason.2017,Bird.2019,Bird.2021,Deason.2021}. Figure~\ref{fig:reflexmodel}  illustrates this by showing the reflex motion signature projected in galactic coordinates in the line-of-sight velocity component. The most significant amount of reflex occurs near the apex and anti-apex locations on the sky in the line-of-sight velocity.

\section{Conclusions}
\label{sec:conclusions}

Using a sample of halo stars, we measured the dipole signal from the motion of halo stars w.r.t the disc. We compare the signal to a suite of MW--LMC models presented here which indicate the origin of the dipole owes to the LMC infall. Following-up on \citetalias{PP21.2021NatAs...5..251P}, we compute the reflex motion parameters as a function of Galactocentric distance. Studying the radial dependence of the reflex motion signal enables more detailed comparison to models of the MW--LMC interaction.

The main results are as follows:

\begin{enumerate}
    \item The measured travel velocity of the disc increases with distance. This agrees with simulations, which predict that the reflex dipole grows with distance and becomes constant at large radii. We find two distinct behaviours. The first is where the response of the halo is adiabatic, reducing significantly the imprint of the dipole. The second is where the magnitude of disc motion is maximal which is the impulsive regime. Current data suggests that the transition occurs over a wide range of radii, centred at $r\simeq 40\kpc$.

    \item Information on the previous trajectory of the LMC is encoded in the direction of the reflex motion. We found that at the largest distance, the apex location points to a location on the sky that is roughly aligned with the previous trajectory of the LMC as found in \citetalias{PP21.2021NatAs...5..251P}.

    \item We find that both the data and simulations show that there is a net inward motion of the halo stars at $r>30~\kpc$ (i.e. compression), which may be due to the deposition of the LMC's bound mass to the inner regions of the MW DM halo. Tests using idealised models show that magnitude of the radial component of the halo bulk motion increases depending on the assumed mass of the LMC. 

    \item By analysing idealised self-consistent models available in the literature (see Table~\ref{tab:models}), we develop a technique to measure the reflex motion parameters ($\vecvtravel$ and bulk motions) directly from simulations. Using this technique, we compare the reflex motion present in the simulations to the data and arrive at the following conclusions (Figures~\ref{fig:apex} and \ref{fig:bulk_motion}). First, in both idealised and self-consistent models, a heavier LMC leads to a larger magnitude of disc motion at all radii. Secondly, none of the simulations reproduce the observed rotation in the halo measured in $\langle \vphi\rangle$, which suggests that the stellar halo may have had a slight prograde rotation prior to the LMC infall. Third, the profiles of $\vecvtravel$ with distance vary least in the magnitude and most in the apex locations. These results call for future simulations that model the reflex motion and the bulk velocities of the stellar halo simultaneously.

    \item We explore the effect of the data on the fitted parameters by using the idealised simulations to test sky coverage effects, binning choice effects and the signal measured for K giants and BHBs independently. We find that using the SDSS footprint is sufficient for the fit to recover the reflex motion parameters computed directly from simulations. We also find that varying the bin locations does not change the trends observed in the data. We find that the fitting results for K giants and BHBs are consistent at most radii, with exception to the 30-40 kpc bin in the apex latitude. 

    \item While self-consistent live $n$-body models are expensive, they are required to fully model the MW--LMC system and in particular to obtain realistic reflex motion signals. Idealised models must have a responsive MW DM halo at the very least, but can still be useful for discovering dynamical mechanisms.
    
    \item We show the all-sky line of sight velocity imprint from reflex motion (Figure~\ref{fig:reflexmodel}), and show that there is a need to correct for the reflex motion of the disc at various radii.
\end{enumerate}

With new large samples of halo stars becoming available in the coming years \citep[e.g.][]{Conroy_2019_H3, Cooper_DESI_MWS, kollmeier2017sdssv}, more precise measurements of the reflex motion parameters will be available. In tandem, the methodology for measuring the parameters may be improved by first transitioning to a bin-free likelihood, and second including higher-order deformations of the halo. Additionally, future models of the MW--LMC interaction will be beneficial for comparison to the data. As the data improves, we can hope for improved modeling schemes that can efficiently search the parameter space of models and orbital histories. Efficient codes for simulating the MW--LMC interaction, such as {\sc exp} \citep{Petersen.exp.2022}, are a promising avenue, with the efficiency to produce high fidelity models. We will present new models in an upcoming work.

\section*{Acknowledgements}

MSP acknowledges funding from a UKRI Stephen Hawking Fellowship. We thank Nico Garavito-Camargo for providing us with a copy of his present-day snapshot. We thank Sophia Lilleengen for comments on a draft of this work. {\sc exp} is maintained by the B-BFE collaboration.  We acknowledge and thank the developers of the following software that was used in this work: NumPy~\citep{Bellazini_ibata_2020}, SciPy~\citep{2020SciPy-NMeth}, IPython~\citep{IPYTHON_PER-GRA:2007}, Matplotlib~\citep{mpl_Hunter:2007}, Jupyter~\citep{community_jupyter_2021}, MultiNest~\citep{Multinest_Feroz_2009} and PyMultiNest~\citep{pymultinest_buchner}.
    
\section*{Data Availability}

The stellar data are available in public catalogues, but our collated version of the data is available upon request. The \citetalias{Erkal.modelrelease.2020} and \citetalias{Vasiliev.secondpassage.2024} models are available with the associated published works. The live \citetalias{Erkal_19_LMC_MASS.10.1093/mnras/stz1371} model, the \citetalias{2020MNRASReflexPP20} models, the \citetalias{Donaldson.2022} model, and the \citetalias{Lilleengen.oc.2023} model are available on reasonable request. The \citetalias{GaravitoCamargo.model.2019} model was obtained directly from Nico Garavito-Camargo.


\bibliographystyle{mnras}
\bibliography{references}

\begin{thebibliography}{}
\makeatletter
\relax
\def\mn@urlcharsother{\let\do\@makeother \do\$\do\&\do\#\do\^\do\_\do\%\do\~}
\def\mn@doi{\begingroup\mn@urlcharsother \@ifnextchar [ {\mn@doi@}
  {\mn@doi@[]}}
\def\mn@doi@[#1]#2{\def\@tempa{#1}\ifx\@tempa\@empty \href
  {http://dx.doi.org/#2} {doi:#2}\else \href {http://dx.doi.org/#2} {#1}\fi
  \endgroup}
\def\mn@eprint#1#2{\mn@eprint@#1:#2::\@nil}
\def\mn@eprint@arXiv#1{\href {http://arxiv.org/abs/#1} {{\tt arXiv:#1}}}
\def\mn@eprint@dblp#1{\href {http://dblp.uni-trier.de/rec/bibtex/#1.xml}
  {dblp:#1}}
\def\mn@eprint@#1:#2:#3:#4\@nil{\def\@tempa {#1}\def\@tempb {#2}\def\@tempc
  {#3}\ifx \@tempc \@empty \let \@tempc \@tempb \let \@tempb \@tempa \fi \ifx
  \@tempb \@empty \def\@tempb {arXiv}\fi \@ifundefined
  {mn@eprint@\@tempb}{\@tempb:\@tempc}{\expandafter \expandafter \csname
  mn@eprint@\@tempb\endcsname \expandafter{\@tempc}}}

\bibitem[\protect\citeauthoryear{{Arora}, {Garavito-Camargo}, {Sanderson},
  {Cunningham}, {Wetzel}, {Panithanpaisal}  \& {Barry}}{{Arora}
  et~al.}{2023}]{Arora.etal.2023}
{Arora} A.,  {Garavito-Camargo} N.,  {Sanderson} R.~E.,  {Cunningham} E.~C.,
  {Wetzel} A.,  {Panithanpaisal} N.,   {Barry} M.,  2023, \mn@doi [arXiv
  e-prints] {10.48550/arXiv.2309.15998}, \href
  {https://ui.adsabs.harvard.edu/abs/2023arXiv230915998A} {p. arXiv:2309.15998}

\bibitem[\protect\citeauthoryear{Bellazzini, Ibata, Malhan, Martin, Famaey  \&
  Thomas}{Bellazzini et~al.}{2020}]{Bellazini_ibata_2020}
Bellazzini M.,  Ibata R.,  Malhan K.,  Martin N.,  Famaey B.,   Thomas G.,
  2020, \mn@doi [\aap] {10.1051/0004-6361/202037621}, 636, A107

\bibitem[\protect\citeauthoryear{Bennett \& Bovy}{Bennett \&
  Bovy}{2018}]{bennet_bovy_19}
Bennett M.,  Bovy J.,  2018, \mn@doi [Monthly Notices of the Royal Astronomical
  Society] {10.1093/mnras/sty2813}, 482, 1417

\bibitem[\protect\citeauthoryear{Besla, Kallivayalil, Hernquist, van~der Marel,
  Cox  \& Kere{\v{s}}}{Besla et~al.}{2010}]{Besla_kalliv_2020_first_infall}
Besla G.,  Kallivayalil N.,  Hernquist L.,  van~der Marel R.,  Cox T.,
  Kere{\v{s}} D.,  2010, \mn@doi [\apjl] {10.1088/2041-8205/721/2/L97}, 721,
  L97

\bibitem[\protect\citeauthoryear{Besla, Kallivayalil, Hernquist, van~der Marel,
  Cox  \& Kere{\v{s}}}{Besla et~al.}{2012}]{Besla_Kalliv_2012}
Besla G.,  Kallivayalil N.,  Hernquist L.,  van~der Marel R.~P.,  Cox T.~J.,
  Kere{\v{s}} D.,  2012, \mn@doi [Monthly Notices of the Royal Astronomical
  Society] {10.1111/j.1365-2966.2012.20466.x}, 421, 2109

\bibitem[\protect\citeauthoryear{{Bird}, {Xue}, {Liu}, {Shen}, {Flynn}  \&
  {Yang}}{{Bird} et~al.}{2019}]{Bird.2019}
{Bird} S.~A.,  {Xue} X.-X.,  {Liu} C.,  {Shen} J.,  {Flynn} C.,   {Yang} C.,
  2019, \mn@doi [\aj] {10.3847/1538-3881/aafd2e}, \href
  {https://ui.adsabs.harvard.edu/abs/2019AJ....157..104B} {157, 104}

\bibitem[\protect\citeauthoryear{{Bird}, {Xue}, {Liu}, {Shen}, {Flynn}, {Yang},
  {Zhao}  \& {Tian}}{{Bird} et~al.}{2021}]{Bird.2021}
{Bird} S.~A.,  {Xue} X.-X.,  {Liu} C.,  {Shen} J.,  {Flynn} C.,  {Yang} C.,
  {Zhao} G.,   {Tian} H.-J.,  2021, \mn@doi [\apj] {10.3847/1538-4357/abfa9e},
  \href {https://ui.adsabs.harvard.edu/abs/2021ApJ...919...66B} {919, 66}

\bibitem[\protect\citeauthoryear{{Bovy}}{{Bovy}}{2015}]{Bovy.galpy.2015}
{Bovy} J.,  2015, \mn@doi [\apjs] {10.1088/0067-0049/216/2/29}, \href
  {https://ui.adsabs.harvard.edu/abs/2015ApJS..216...29B} {216, 29}

\bibitem[\protect\citeauthoryear{Bovy, Bahmanyar, Fritz  \& Kallivayalil}{Bovy
  et~al.}{2016}]{Bovy_2016}
Bovy J.,  Bahmanyar A.,  Fritz T.~K.,   Kallivayalil N.,  2016, \mn@doi [The
  Astrophysical Journal] {10.3847/1538-4357/833/1/31}, 833, 31

\bibitem[\protect\citeauthoryear{{Brooks}, {Sanders}, {Lilleengen}, {Petersen}
  \& {Pontzen}}{{Brooks} et~al.}{2024}]{Brooks.etal.2024}
{Brooks} R. A.~N.,  {Sanders} J.~L.,  {Lilleengen} S.,  {Petersen} M.~S.,
  {Pontzen} A.,  2024, \mn@doi [arXiv e-prints] {10.48550/arXiv.2401.11990},
  \href {https://ui.adsabs.harvard.edu/abs/2024arXiv240111990B} {p.
  arXiv:2401.11990}

\bibitem[\protect\citeauthoryear{{Buchner} et~al.,}{{Buchner}
  et~al.}{2014}]{pymultinest_buchner}
{Buchner} J.,  et~al., 2014, \mn@doi [\aap] {10.1051/0004-6361/201322971},
  \href {https://ui.adsabs.harvard.edu/abs/2014A&A...564A.125B} {564, A125}

\bibitem[\protect\citeauthoryear{Community}{Community}{2021}]{community_jupyter_2021}
Community E.~B.,  2021, Jupyter {Book}, \mn@doi{10.5281/zenodo.4539666}, \url
  {https://doi.org/10.5281/zenodo.4539666}

\bibitem[\protect\citeauthoryear{Conroy et~al.,}{Conroy
  et~al.}{2019}]{Conroy_2019_H3}
Conroy C.,  et~al., 2019, \mn@doi [The Astrophysical Journal]
  {10.3847/1538-4357/ab38b8}, 883, 107

\bibitem[\protect\citeauthoryear{{Conroy}, {Naidu}, {Garavito-Camargo},
  {Besla}, {Zaritsky}, {Bonaca}  \& {Johnson}}{{Conroy}
  et~al.}{2021}]{conroy.2021.response}
{Conroy} C.,  {Naidu} R.~P.,  {Garavito-Camargo} N.,  {Besla} G.,  {Zaritsky}
  D.,  {Bonaca} A.,   {Johnson} B.~D.,  2021, \mn@doi [\nat]
  {10.1038/s41586-021-03385-7}, \href
  {https://ui.adsabs.harvard.edu/abs/2021Natur.592..534C} {592, 534}

\bibitem[\protect\citeauthoryear{{Cooper} et~al.,}{{Cooper}
  et~al.}{2023}]{Cooper_DESI_MWS}
{Cooper} A.~P.,  et~al., 2023, \mn@doi [\apj] {10.3847/1538-4357/acb3c0}, \href
  {https://ui.adsabs.harvard.edu/abs/2023ApJ...947...37C} {947, 37}

\bibitem[\protect\citeauthoryear{{Cunningham} et~al.,}{{Cunningham}
  et~al.}{2020}]{cunningham.2020.responseharmonic}
{Cunningham} E.~C.,  et~al., 2020, \mn@doi [\apj] {10.3847/1538-4357/ab9b88},
  \href {https://ui.adsabs.harvard.edu/abs/2020ApJ...898....4C} {898, 4}

\bibitem[\protect\citeauthoryear{{Deason}, {Belokurov}, {Koposov}, {G{\'o}mez},
  {Grand}, {Marinacci}  \& {Pakmor}}{{Deason} et~al.}{2017}]{Deason.2017}
{Deason} A.~J.,  {Belokurov} V.,  {Koposov} S.~E.,  {G{\'o}mez} F.~A.,  {Grand}
  R.~J.,  {Marinacci} F.,   {Pakmor} R.,  2017, \mn@doi [\mnras]
  {10.1093/mnras/stx1301}, \href
  {https://ui.adsabs.harvard.edu/abs/2017MNRAS.470.1259D} {470, 1259}

\bibitem[\protect\citeauthoryear{{Deason} et~al.,}{{Deason}
  et~al.}{2021}]{Deason.2021}
{Deason} A.~J.,  et~al., 2021, \mn@doi [\mnras] {10.1093/mnras/staa3984}, \href
  {https://ui.adsabs.harvard.edu/abs/2021MNRAS.501.5964D} {501, 5964}

\bibitem[\protect\citeauthoryear{{Donaldson}, {Petersen}  \&
  {Pe{\~n}arrubia}}{{Donaldson} et~al.}{2022}]{Donaldson.2022}
{Donaldson} K.,  {Petersen} M.~S.,   {Pe{\~n}arrubia} J.,  2022, \mn@doi
  [\mnras] {10.1093/mnrasl/slac031}, \href
  {https://ui.adsabs.harvard.edu/abs/2022MNRAS.513L..46D} {513, 46}

\bibitem[\protect\citeauthoryear{{Eilers}, {Hogg}, {Rix}  \& {Ness}}{{Eilers}
  et~al.}{2019}]{Eilers_19_mwcircvel}
{Eilers} A.-C.,  {Hogg} D.~W.,  {Rix} H.-W.,   {Ness} M.~K.,  2019, \mn@doi
  [\apj] {10.3847/1538-4357/aaf648}, \href
  {https://ui.adsabs.harvard.edu/abs/2019ApJ...871..120E} {871, 120}

\bibitem[\protect\citeauthoryear{Erkal et~al.,}{Erkal
  et~al.}{2019}]{Erkal_19_LMC_MASS.10.1093/mnras/stz1371}
Erkal D.,  et~al., 2019, \mn@doi [Monthly Notices of the Royal Astronomical
  Society] {10.1093/mnras/stz1371}, 487, 2685

\bibitem[\protect\citeauthoryear{{Erkal}, {Belokurov}  \& {Parkin}}{{Erkal}
  et~al.}{2020}]{Erkal.modelrelease.2020}
{Erkal} D.,  {Belokurov} V.~A.,   {Parkin} D.~L.,  2020, \mn@doi [\mnras]
  {10.1093/mnras/staa2840}, \href
  {https://ui.adsabs.harvard.edu/abs/2020MNRAS.498.5574E} {498, 5574}

\bibitem[\protect\citeauthoryear{{Erkal} et~al.,}{{Erkal}
  et~al.}{2021}]{Erkal.2021.SLOSHING}
{Erkal} D.,  et~al., 2021, \mn@doi [\mnras] {10.1093/mnras/stab1828}, \href
  {https://ui.adsabs.harvard.edu/abs/2021MNRAS.506.2677E} {506, 2677}

\bibitem[\protect\citeauthoryear{Fardal, van der Marel, Law, Sohn, Sesar,
  Hernitschek  \& Rix}{Fardal et~al.}{2018}]{Fardal_2005}
Fardal M.~A.,  van der Marel R.~P.,  Law D.~R.,  Sohn S.~T.,  Sesar B.,
  Hernitschek N.,   Rix H.-W.,  2018, \mn@doi [Monthly Notices of the Royal
  Astronomical Society] {10.1093/mnras/sty3428}, 483, 4724

\bibitem[\protect\citeauthoryear{Feroz \& Hobson}{Feroz \&
  Hobson}{2008}]{Feroz_Hobson_Multinest.10.1111/j.1365-2966.2007.12353.x}
Feroz F.,  Hobson M.~P.,  2008, \mn@doi [Monthly Notices of the Royal
  Astronomical Society] {10.1111/j.1365-2966.2007.12353.x}, 384, 449

\bibitem[\protect\citeauthoryear{Feroz, Hobson  \& Bridges}{Feroz
  et~al.}{2009}]{Multinest_Feroz_2009}
Feroz F.,  Hobson M.~P.,   Bridges M.,  2009, \mn@doi [Monthly Notices of the
  Royal Astronomical Society] {10.1111/j.1365-2966.2009.14548.x}, 398,
  1601–1614

\bibitem[\protect\citeauthoryear{{GRAVITY Collaboration} et~al.,}{{GRAVITY
  Collaboration} et~al.}{2019}]{Gravity_collab_19}
{GRAVITY Collaboration} et~al., 2019, \mn@doi [\aap]
  {10.1051/0004-6361/201935656}, 625, L10

\bibitem[\protect\citeauthoryear{{Garavito-Camargo}, {Besla}, {Laporte},
  {Johnston}, {G{\'o}mez}  \& {Watkins}}{{Garavito-Camargo}
  et~al.}{2019}]{GaravitoCamargo.model.2019}
{Garavito-Camargo} N.,  {Besla} G.,  {Laporte} C. F.~P.,  {Johnston} K.~V.,
  {G{\'o}mez} F.~A.,   {Watkins} L.~L.,  2019, \mn@doi [\apj]
  {10.3847/1538-4357/ab32eb}, \href
  {https://ui.adsabs.harvard.edu/abs/2019ApJ...884...51G} {884, 51}

\bibitem[\protect\citeauthoryear{{Garavito-Camargo}, {Besla}, {Laporte},
  {Price-Whelan}, {Cunningham}, {Johnston}, {Weinberg}  \&
  {G{\'o}mez}}{{Garavito-Camargo} et~al.}{2021}]{GaravitoCamargo.bfe.2021}
{Garavito-Camargo} N.,  {Besla} G.,  {Laporte} C. F.~P.,  {Price-Whelan} A.~M.,
   {Cunningham} E.~C.,  {Johnston} K.~V.,  {Weinberg} M.,   {G{\'o}mez} F.~A.,
  2021, \mn@doi [\apj] {10.3847/1538-4357/ac0b44}, \href
  {https://ui.adsabs.harvard.edu/abs/2021ApJ...919..109G} {919, 109}

\bibitem[\protect\citeauthoryear{G{\'{o}}mez, Besla, Carpintero, Villalobos,
  O'Shea  \& Bell}{G{\'{o}}mez et~al.}{2015}]{G_mez_2015}
G{\'{o}}mez F.~A.,  Besla G.,  Carpintero D.~D.,  Villalobos {\'{A}}.,  O'Shea
  B.~W.,   Bell E.~F.,  2015, \mn@doi [The Astrophysical Journal]
  {10.1088/0004-637x/802/2/128}, 802, 128

\bibitem[\protect\citeauthoryear{{Hernquist}}{{Hernquist}}{1990}]{Hernquist.1990}
{Hernquist} L.,  1990, \mn@doi [\apj] {10.1086/168845}, \href
  {https://ui.adsabs.harvard.edu/abs/1990ApJ...356..359H} {356, 359}

\bibitem[\protect\citeauthoryear{Hunter}{Hunter}{2007}]{mpl_Hunter:2007}
Hunter J.~D.,  2007, \mn@doi [Computing in Science \& Engineering]
  {10.1109/MCSE.2007.55}, 9, 90

\bibitem[\protect\citeauthoryear{Johnston, Law  \& Majewski}{Johnston
  et~al.}{2005}]{Johnston_2005}
Johnston K.~V.,  Law D.~R.,   Majewski S.~R.,  2005, \mn@doi [The Astrophysical
  Journal] {10.1086/426777}, 619, 800

\bibitem[\protect\citeauthoryear{Kollmeier et~al.,}{Kollmeier
  et~al.}{2017}]{kollmeier2017sdssv}
Kollmeier J.~A.,  et~al., 2017, SDSS-V: Pioneering Panoptic Spectroscopy
  (\mn@eprint {arXiv} {1711.03234})

\bibitem[\protect\citeauthoryear{Koposov et~al.,}{Koposov
  et~al.}{2023}]{ORPHAN_KOPOSOV.2023MNRAS.521.4936K}
Koposov S.~E.,  et~al., 2023, \mn@doi [\mnras] {10.1093/mnras/stad551}, 521,
  4936

\bibitem[\protect\citeauthoryear{{Lancaster}, {Koposov}, {Belokurov}, {Evans}
  \& {Deason}}{{Lancaster} et~al.}{2019}]{Lancaster.2019}
{Lancaster} L.,  {Koposov} S.~E.,  {Belokurov} V.,  {Evans} N.~W.,   {Deason}
  A.~J.,  2019, \mn@doi [\mnras] {10.1093/mnras/stz853}, \href
  {https://ui.adsabs.harvard.edu/abs/2019MNRAS.486..378L} {486, 378}

\bibitem[\protect\citeauthoryear{Law \& Majewski}{Law \&
  Majewski}{2010}]{Law_2010.halo_sgr_oblate}
Law D.~R.,  Majewski S.~R.,  2010, \mn@doi [The Astrophysical Journal]
  {10.1088/0004-637X/714/1/229}, 714, 229

\bibitem[\protect\citeauthoryear{{Lilleengen} et~al.,}{{Lilleengen}
  et~al.}{2023}]{Lilleengen.oc.2023}
{Lilleengen} S.,  et~al., 2023, \mn@doi [\mnras] {10.1093/mnras/stac3108},
  \href {https://ui.adsabs.harvard.edu/abs/2023MNRAS.518..774L} {518, 774}

\bibitem[\protect\citeauthoryear{{Majewski}, {Skrutskie}, {Weinberg}  \&
  {Ostheimer}}{{Majewski} et~al.}{2003}]{Majewski.2003.2MASS.SGR}
{Majewski} S.~R.,  {Skrutskie} M.~F.,  {Weinberg} M.~D.,   {Ostheimer} J.~C.,
  2003, \mn@doi [\apj] {10.1086/379504}, \href
  {https://ui.adsabs.harvard.edu/abs/2003ApJ...599.1082M} {599, 1082}

\bibitem[\protect\citeauthoryear{{Navarro}, {Frenk}  \& {White}}{{Navarro}
  et~al.}{1997}]{Navarro.Frenk.White.1997}
{Navarro} J.~F.,  {Frenk} C.~S.,   {White} S. D.~M.,  1997, \mn@doi [\apj]
  {10.1086/304888}, \href
  {https://ui.adsabs.harvard.edu/abs/1997ApJ...490..493N} {490, 493}

\bibitem[\protect\citeauthoryear{{Pe{\~n}arrubia} \&
  {Petersen}}{{Pe{\~n}arrubia} \& {Petersen}}{2021}]{SGR.GMM.PP21}
{Pe{\~n}arrubia} J.,  {Petersen} M.~S.,  2021, \mn@doi [\mnras]
  {10.1093/mnrasl/slab090}, \href
  {https://ui.adsabs.harvard.edu/abs/2021MNRAS.508L..26P} {508, L26}

\bibitem[\protect\citeauthoryear{Pe{\~{n}}arrubia, G{\'{o}}mez, Besla, Erkal
  \& Ma}{Pe{\~{n}}arrubia et~al.}{2016}]{LMCTIMING_JP.2016MNRAS.456L..54P}
Pe{\~{n}}arrubia J.,  G{\'{o}}mez F.~A.,  Besla G.,  Erkal D.,   Ma Y.-Z.,
  2016, \mn@doi [\mnras] {10.1093/mnrasl/slv160}, 456, L54

\bibitem[\protect\citeauthoryear{P\'erez \& Granger}{P\'erez \&
  Granger}{2007}]{IPYTHON_PER-GRA:2007}
P\'erez F.,  Granger B.~E.,  2007, \mn@doi [Computing in Science and
  Engineering] {10.1109/MCSE.2007.53}, 9, 21

\bibitem[\protect\citeauthoryear{Petersen \& Pe{\~{n}}arrubia}{Petersen \&
  Pe{\~{n}}arrubia}{2020}]{2020MNRASReflexPP20}
Petersen M.~S.,  Pe{\~{n}}arrubia J.,  2020, \mn@doi [\mnras]
  {10.1093/mnrasl/slaa029}, 494, L11

\bibitem[\protect\citeauthoryear{Petersen \& Pe{\~{n}}arrubia}{Petersen \&
  Pe{\~{n}}arrubia}{2021}]{PP21.2021NatAs...5..251P}
Petersen M.~S.,  Pe{\~{n}}arrubia J.,  2021, \mn@doi [Nature Astronomy]
  {10.1038/s41550-020-01254-3}, 5, 251

\bibitem[\protect\citeauthoryear{{Petersen}, {Weinberg}  \& {Katz}}{{Petersen}
  et~al.}{2022a}]{Petersen.exp.2022}
{Petersen} M.~S.,  {Weinberg} M.~D.,   {Katz} N.,  2022a, \mn@doi [\mnras]
  {10.1093/mnras/stab3639}, \href
  {https://ui.adsabs.harvard.edu/abs/2022MNRAS.510.6201P} {510, 6201}

\bibitem[\protect\citeauthoryear{{Petersen}, {Pe{\~n}arrubia}  \&
  {Jones}}{{Petersen} et~al.}{2022b}]{PPJ.2022}
{Petersen} M.~S.,  {Pe{\~n}arrubia} J.,   {Jones} E.,  2022b, \mn@doi [\mnras]
  {10.1093/mnras/stac1429}, \href
  {https://ui.adsabs.harvard.edu/abs/2022MNRAS.514.1266P} {514, 1266}

\bibitem[\protect\citeauthoryear{{Plummer}}{{Plummer}}{1911}]{Plummer.1911}
{Plummer} H.~C.,  1911, \mn@doi [\mnras] {10.1093/mnras/71.5.460}, \href
  {https://ui.adsabs.harvard.edu/abs/1911MNRAS..71..460P} {71, 460}

\bibitem[\protect\citeauthoryear{{Rozier}, {Famaey}, {Siebert}, {Monari},
  {Pichon}  \& {Ibata}}{{Rozier} et~al.}{2022}]{Rozier.etal.2022}
{Rozier} S.,  {Famaey} B.,  {Siebert} A.,  {Monari} G.,  {Pichon} C.,   {Ibata}
  R.,  2022, \mn@doi [\apj] {10.3847/1538-4357/ac7139}, \href
  {https://ui.adsabs.harvard.edu/abs/2022ApJ...933..113R} {933, 113}

\bibitem[\protect\citeauthoryear{Schönrich, Binney  \& Dehnen}{Schönrich
  et~al.}{2010}]{schonrich_binney_dehnen_10}
Schönrich R.,  Binney J.,   Dehnen W.,  2010, \mn@doi [Monthly Notices of the
  Royal Astronomical Society] {10.1111/j.1365-2966.2010.16253.x}, 403, 1829

\bibitem[\protect\citeauthoryear{{Shipp} et~al.,}{{Shipp}
  et~al.}{2019}]{Shipp.etal.2019}
{Shipp} N.,  et~al., 2019, \mn@doi [\apj] {10.3847/1538-4357/ab44bf}, \href
  {https://ui.adsabs.harvard.edu/abs/2019ApJ...885....3S} {885, 3}

\bibitem[\protect\citeauthoryear{{Vasiliev}}{{Vasiliev}}{2024}]{Vasiliev.secondpassage.2024}
{Vasiliev} E.,  2024, \mn@doi [\mnras] {10.1093/mnras/stad2612}, \href
  {https://ui.adsabs.harvard.edu/abs/2024MNRAS.527..437V} {527, 437}

\bibitem[\protect\citeauthoryear{Virtanen et~al.,}{Virtanen
  et~al.}{2020}]{2020SciPy-NMeth}
Virtanen P.,  et~al., 2020, \mn@doi [Nature Methods]
  {10.1038/s41592-019-0686-2}, \href {https://rdcu.be/b08Wh} {17, 261}

\bibitem[\protect\citeauthoryear{Yanny et~al.,}{Yanny
  et~al.}{2009}]{SEGUE.Yanny_2009}
Yanny B.,  et~al., 2009, \mn@doi [The Astronomical Journal]
  {10.1088/0004-6256/137/5/4377}, 137, 4377

\bibitem[\protect\citeauthoryear{{van der Marel} \& {Kallivayalil}}{{van der
  Marel} \& {Kallivayalil}}{2014}]{vanderMarel.2014}
{van der Marel} R.~P.,  {Kallivayalil} N.,  2014, \mn@doi [\apj]
  {10.1088/0004-637X/781/2/121}, \href
  {https://ui.adsabs.harvard.edu/abs/2014ApJ...781..121V} {781, 121}

\makeatother
\end{thebibliography}


\appendix

\section{Sky Coverage Effects}
\label{appendix:b}

In this Appendix, we test the effects of sky coverage on the fitted model parameters by constructing three samples from the $n$-body catalogues: (i) an all-sky sample, (ii) a SDSS-footprint-only sample, and (iii) a sample with stars only in the northern galactic hemisphere ($b>0^{\circ}$). 

\begin{figure}
    \includegraphics[width=\columnwidth]{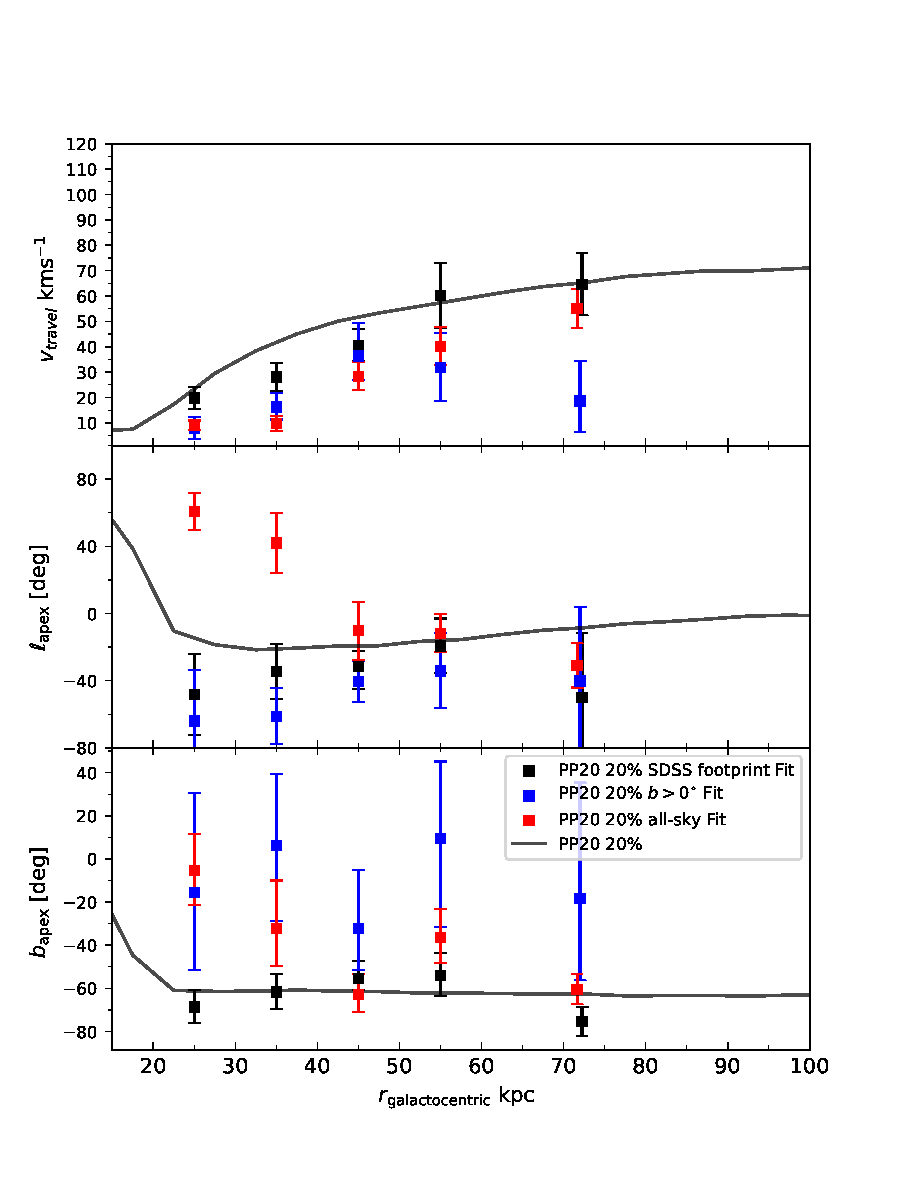}
    \caption{ \label{fig:A2-skyselction-reflex} Sky-coverage tests on a mock sample for recovery of the dipole parameters. The measured values of the travel velocity and apex directions for the \citetalias{2020MNRASReflexPP20} 20\% model versus the median galactocentric distance of stars in each bin.  Top panel: fitted travel velocity in each bin with uncertainties given as the standard deviation of the posterior chains of the parameters. Middle panel: Measured $\lapex$ values for stars in each bin, note that we restrict the apex longitude to be between $100^{\circ}$ and $-80^{\circ}$. Bottom panel: Measured $\bapex$ values for stars in each bin, where the range of latitude angles are limited to $-90^{\circ}$ and $60^{\circ}$ In each panel, we plot corresponding measurement from the simulation (see text for details in calculating the simulation curves). Colours of model curves and points correspond to the sky coverage used in the test, where the colours red, blue, and black correspond to samples with all-sky, $b>0^{\circ}$ and SDSS footprint tests, respectively. Error bars indicate the 1$\sigma$ width of the posterior distribution for each parameter.}
\end{figure}

\begin{figure}
    \includegraphics[width=\columnwidth]{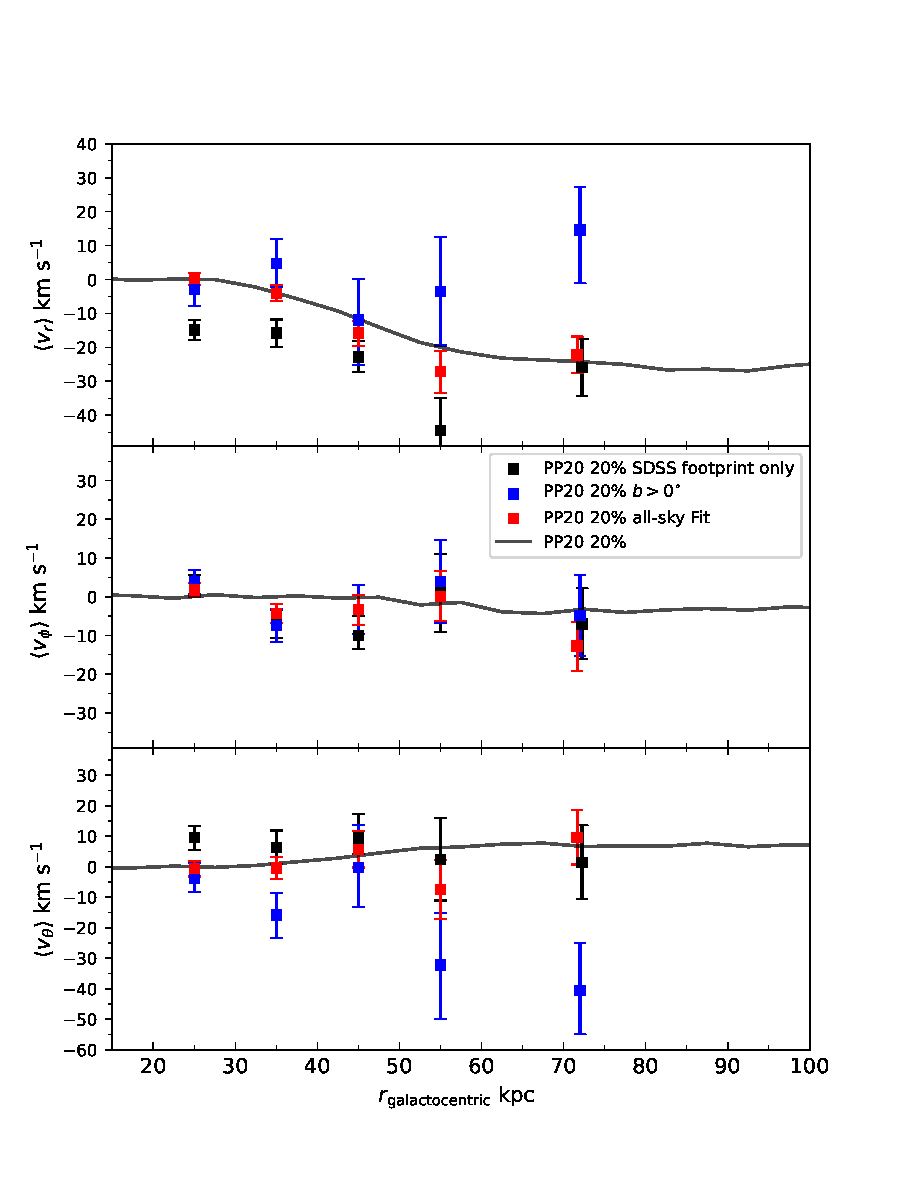}
    \caption{The measured halo bulk motion parameters as a function of galactocentric radius. Top panel: Mean halo motion in the radial direction.This figure shows the sky-coverage tests on the mock sample in the bulk motion parameters. Middle panel: Mean halo motion in the azimuthal direction(cylindrical rotation). Bottom Panel: mean halo motion in the polar direction. In each panel, we plot corresponding measurements from the set of simulations, which have been reflex corrected. Colours of model curves and points correspond to the sky coverage used in the test, where the colours red, blue, and black correspond to samples with all-sky, $b>0^{\circ}$ and SDSS footprint tests, respectively. Error bars indicate the 1$\sigma$ width of the posterior distribution for each parameter.}
    \label{fig:A2-skyselction-bulkmotions}
\end{figure}

Figures~\ref{fig:A2-skyselction-reflex} and \ref{fig:A2-skyselction-bulkmotions} show the result of the fits. In terms of the travel velocity, the north-only fits are not in agreement with either the `true' underlying travel velocity (solid line) or the SDSS fit points, where between 20 and 40~\kpc they are $>1\sigma$ apart. The north-only fits underestimate the travel velocity at all radii. Also, in the bottom panel of Figure \ref{fig:A2-skyselction-reflex}, we find that information about the apex latitude $\bapex$ is completely lost, as the north-only fit values are all consistent zero at all radii. This is not unexpected, as by definition, the apex direction is defined as the location of the stars with the highest travel velocity, which in the case of these simulations, lies in the region of $b<$0$^{\circ}$. 

Furthermore, the fit shows sensitivity to the footprint in which the data lie and $\langle\vr\rangle$ can be shifted towards more negative values as a result. In Figure~\ref{fig:A2-skyselction-bulkmotions} we find that using only the SDSS footprint leads to a difference of approximately 10 kms$^{-1}$ in $\langle \vr\rangle$, while $\langle\vphi\rangle$ and $\langle\vtheta\rangle$ remain roughly consistent. Furthermore, using the all-sky data to fit the bulk motion parameters yields results that are consistent with the underlying mean spherical velocities of the shells of stars from simulation.

To compare the fitted results to the parameter values measured from the simulation in the figures, we include curves from the 20\% LMC model from \citetalias{2020MNRASReflexPP20}. We show the tracks in Figures~\ref{fig:A2-skyselction-reflex} and \ref{fig:A2-skyselction-bulkmotions}. The key result in this Appendix is the difference between the different fit points at a given radius, and the difference between the fit points and the model line. We find that fitting data within the SDSS footprint provides the best match to the $n$-body models.
    
\section{Separate K Giants and BHB fits}
\label{appendix:c}

In order to test biases arising from the choice of stellar halo tracers we fit the K giants and BHBs separately in this Appendix. To facilitate a comparison with other fits in this work, we keep the bins the same, with the exception of one fewer bin for the BHBs, as the sample is much smaller at 50+ kpc.
    
\begin{figure}
    \includegraphics[width=\columnwidth]{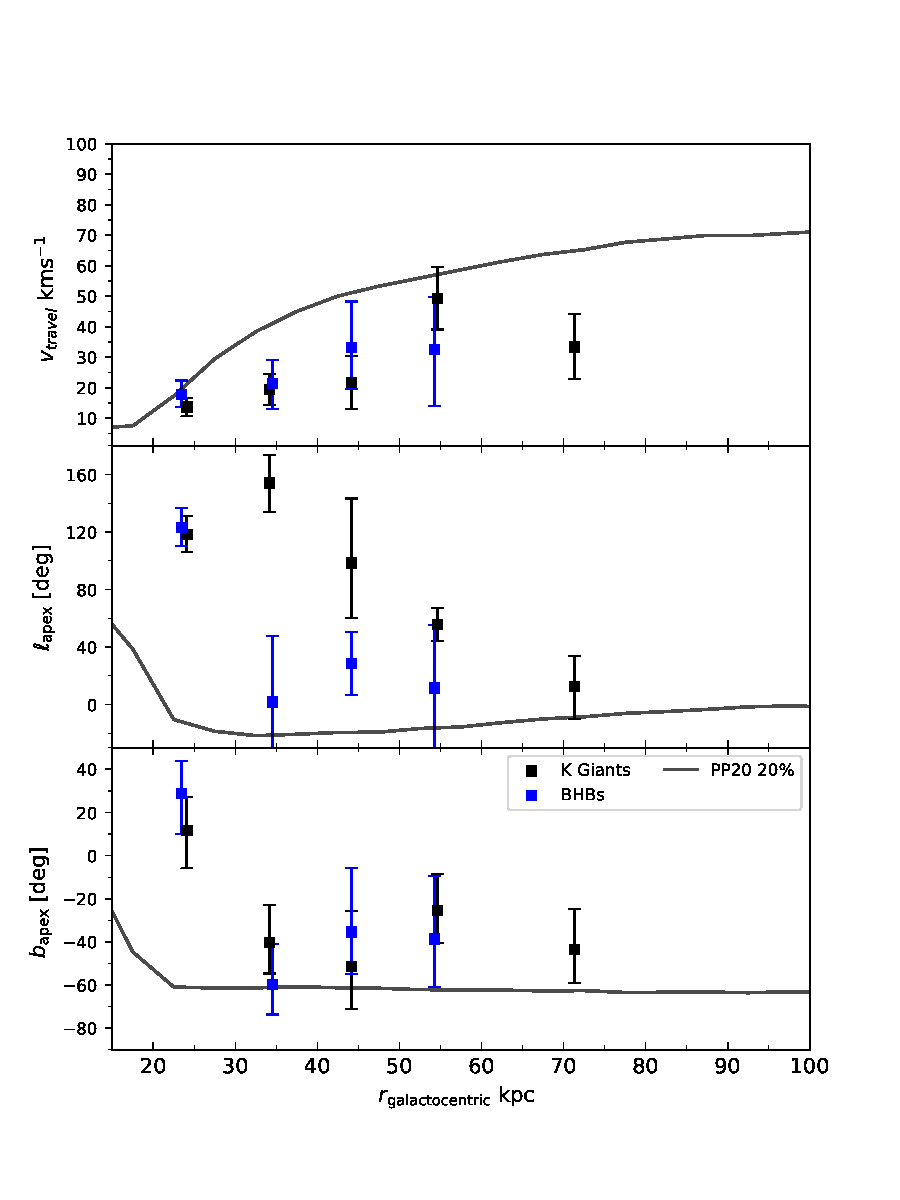}
    \caption{The measured values of the travel velocity and apex directions for the samples of the K giants and BHBs separately, versus the median galactocentric distance of stars in each bin. Top panel: fitted travel velocity in each bin with uncertainties given as the standard deviation of the posterior chains of the parameters. Middle panel: Measured $\lapex$ values for stars in each bin, note that we restrict the apex longitude to be between $180^{\circ}$ and $-20^{\circ}$. Bottom panel: Measured $\bapex$ values for stars in each bin, where the range of latitude angles are limited to $-90^{\circ}$ and $60^{\circ}$ In each panel, we plot the measurement from the \citetalias{2020MNRASReflexPP20} simulation as a visual guide. Colours of the points correspond to the star type used in the test, where the colours blue and black correspond to BHBs and K giants respectively. Error bars indicate the 1$\sigma$ width of the posterior distribution for each parameter.}
    \label{fig:C1-KGvBHB}
\end{figure}

\begin{figure}
    \includegraphics[width=\columnwidth]{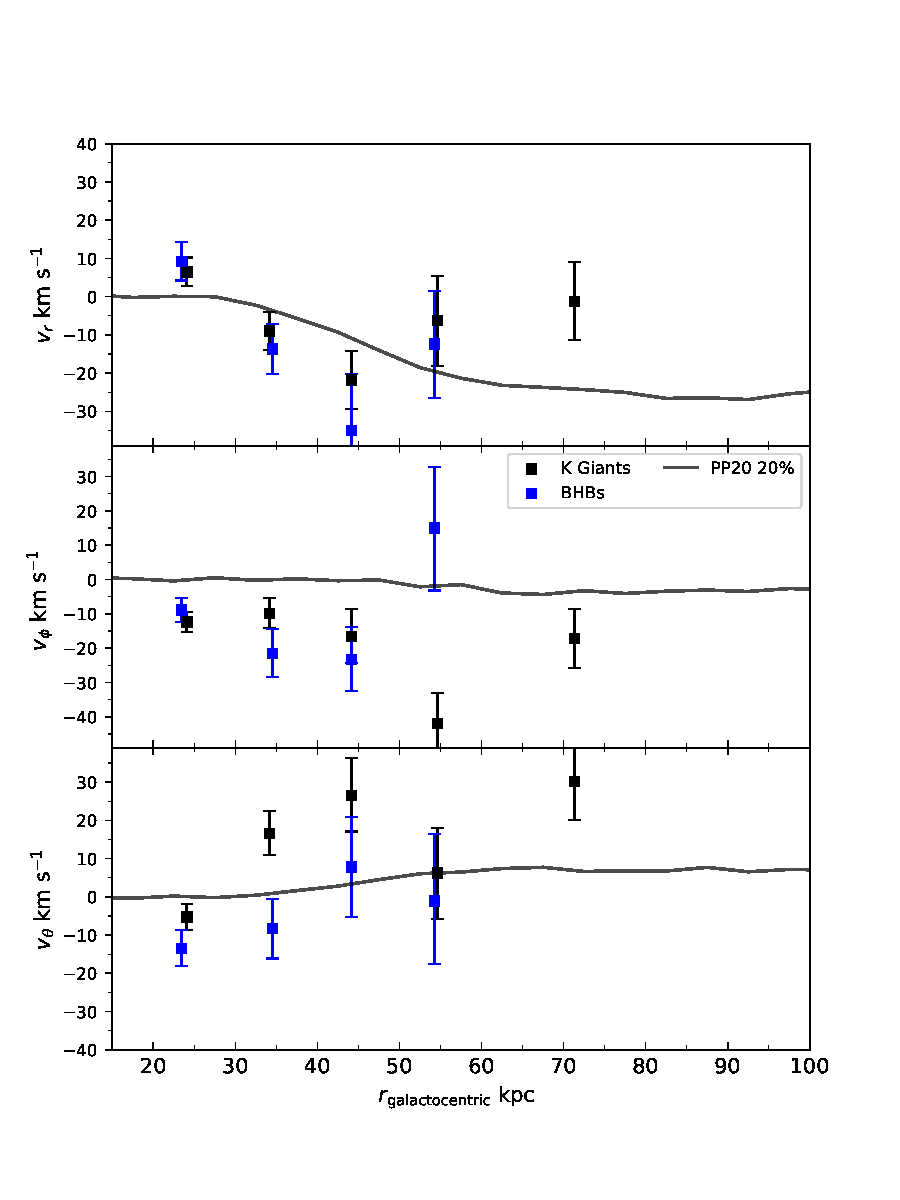}
    \caption{The measured halo bulk motion parameters as a function of galactocentric radius for the K giant and BHBs tests. Top panel: Mean halo motion in the radial direction.This figure shows measured results for the separate BHB or K giants in the bulk motion parameters. Middle panel: Mean halo motion in the azimuthal direction(cylindrical rotation). Bottom Panel: mean halo motion in the polar direction. In each panel, we plot the measurement from the \citetalias{2020MNRASReflexPP20} simulation as a visual guide. Colours of the points correspond to the star type used in the test, where the colours blue and black correspond to BHBs and K giants respectively. Error bars indicate the 1$\sigma$ width of the posterior distribution for each parameter.}
    \label{fig:fig:C2-KGvBHB}
\end{figure}

Figures~\ref{fig:C1-KGvBHB} and \ref{fig:fig:C2-KGvBHB} show the results of the fits. In the case of the reflex motion parameters, there is good agreement between the K Giants and BHBs for all three parameters. In the case of the apex longitude $\lapex$, the BHB fit finds a significantly different result in the 30-40 kpc bin. This could be due to the presence of substructure such as streams. The stream scenario is supported by the result of $\langle v_{\theta}\rangle $ fits at that distance. The K Giants have $\sim$15 kms$^{-1}$, while the BHBs have approximately -10 \kms. Bulk polar motion of a subset of stars in the bin will drastically change the mean the bulk halo motion parameters if a stream, or comoving group of stars exists in that bin. All other bulk motion parameters are in agreement, with exception to the 50-60 kpc bin azimuthal velocity $\langle v_{\phi}\rangle$. We expect this case to also be caused by the presence of substructure, where in that bin the comoving group of stars is deviating the mean halo motion significantly. The travel velocity signal at that radius is underestimated compared to the K giants, and does not display the same increasing $\vtravel$ signal as the K giants, however the BHB fits at radii $>$40kpc all have relatively large error bars due to the small number of stars in those bins ($<$ 100 stars). 

As in Appendix~\ref{appendix:b}, to guide the eye between different figures, we include curves from the 20\% LMC model from \citetalias{2020MNRASReflexPP20}. While we show the tracks in Figures~\ref{fig:A2-skyselction-reflex} and \ref{fig:A2-skyselction-bulkmotions}, the curves should not be interpreted as targets: the key result in this appendix is the difference between the different fit points at a given radius.

 \section{Model fits tested on Mock Data}
\label{sec:mockfits}

In this Appendix, we test the reflex motion model in Section~\ref{sec:methods} using the \citetalias{PP21.2021NatAs...5..251P} models. In the \citetalias[Supplementary Information Section of][]{PP21.2021NatAs...5..251P}, an extensive description of the model parameters, live $n$-body models, LMC trajectory and catalog creation was presented\footnote{In the models as run with {\sc exp}, the LMC is in the wrong present-day position by 90$^\circ$. In \citetalias{PP21.2021NatAs...5..251P}, the coordinate system was rotated to more closely match the present-day location (a transformation that we are free to make, as it only applies a linear offset to $\lapex$). In these Appendices, we choose not to rotate the coordinate system, to more readily match the simulation snapshots offered in the Data Availability section.}. In this work we perform additional analysis on testing the reflex motion model against the mock catalogues generated from those simulations. We bin the mock catalogues in similar bins as the data, and fit all three \citetalias{PP21.2021NatAs...5..251P} models, to verify the fits are able to recover the change in parameters when varying the mass of the LMC (using the three \citetalias{PP21.2021NatAs...5..251P} models with MW--LMC mass ratios of 0.1, 0.2 and 0.3). We bin the mock catalogs between 20 and 60 kpc with bins widths of 10 kpc, the final bin at 60+ kpc contains all stars with distances $r>60$ kpc. The total number of stars in the mocks is limited to 5000 to mimic the numbers of stars in the real dataset. We fit the reflex motion model to the binned mock data.

\begin{figure}
    \includegraphics[width=\columnwidth]{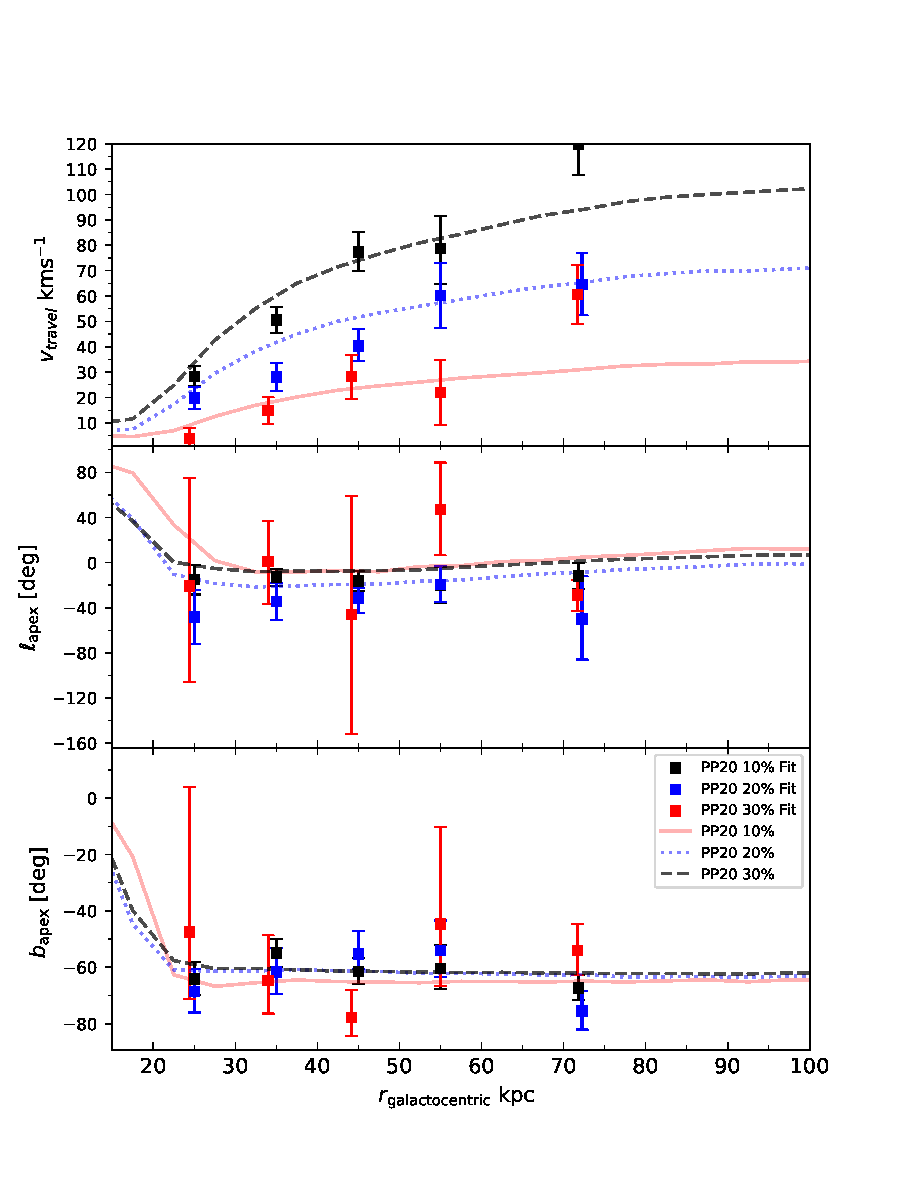}

    \caption{\label{fig:A1-fitverify-reflex} The measured values of the travel velocity and apex directions for the \citetalias{2020MNRASReflexPP20} models versus the median galactocentric distance of stars in each bin. Top panel: fitted travel velocity in each bin with uncertainties given as the standard deviation of the posterior chains of the parameters. Middle panel: Measured $\lapex$ values for stars in each bin, note that we restrict the apex longitude to be between $100^{\circ}$ and $-160^{\circ}$. Bottom panel: Measured $\bapex$ values for stars in each bin, where the range of latitude angles are limited to $-90^{\circ}$ and $15^{\circ}$ In each panel, we plot corresponding measurements from the set of simulations (see Section~\ref{subsec:measuringdipolesim} for details in calculating the simulation curves). Colours of model curves correspond to the ratio of initial LMC mass to MW mass, where the colours red, blue, and black correspond to the \citetalias{2020MNRASReflexPP20} 10\%, 20\% and 30\% models. Error bars indicate the 1$\sigma$ width of the posterior distribution for each parameter.
    }
\end{figure}

\begin{figure}
    \includegraphics[width=\columnwidth]{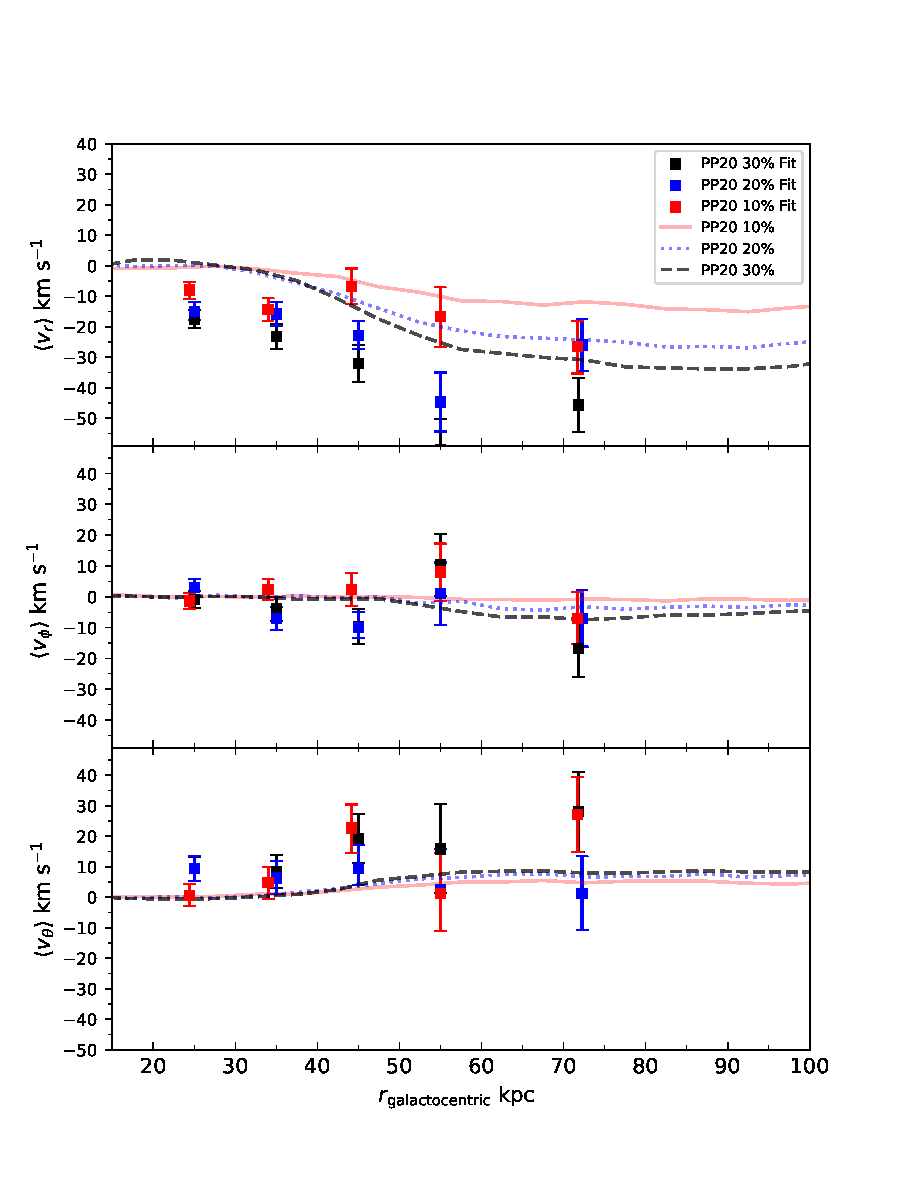}
    \caption{The measured halo bulk motion parameters as a function of galactocentric radius. Top panel: Mean halo motion in the radial direction. Middle panel: Mean halo motion in the azimuthal direction(cylindrical rotation). Bottom Panel: mean halo motion in the polar direction. In each panel, we plot corresponding measurements from the \citetalias{2020MNRASReflexPP20} set of simulations, which have been reflex corrected. Error bars indicate the 1$\sigma$ width of the posterior distribution for each parameter.}
    \label{fig:A1-fitveryfiy-bulkmotions}
\end{figure}

Figures~\ref{fig:A1-fitverify-reflex} and \ref{fig:A1-fitveryfiy-bulkmotions} show the fitted $\vecvtravel$ and bulk motion parameters. At most radii, the fit is $1\sigma$ consistent with the model lines computed directly from the simulations. In the lowest mass \citetalias{2020MNRASReflexPP20} 10\% model, the uncertainties are large due to the weak reflex motion signal, while in the bulk motions, the nonzero radial velocity signal is modestly overestimated at all radii. 

\section{Bin Width Variations}
\label{appendix:d}

\begin{figure}
    \includegraphics[width=\columnwidth]{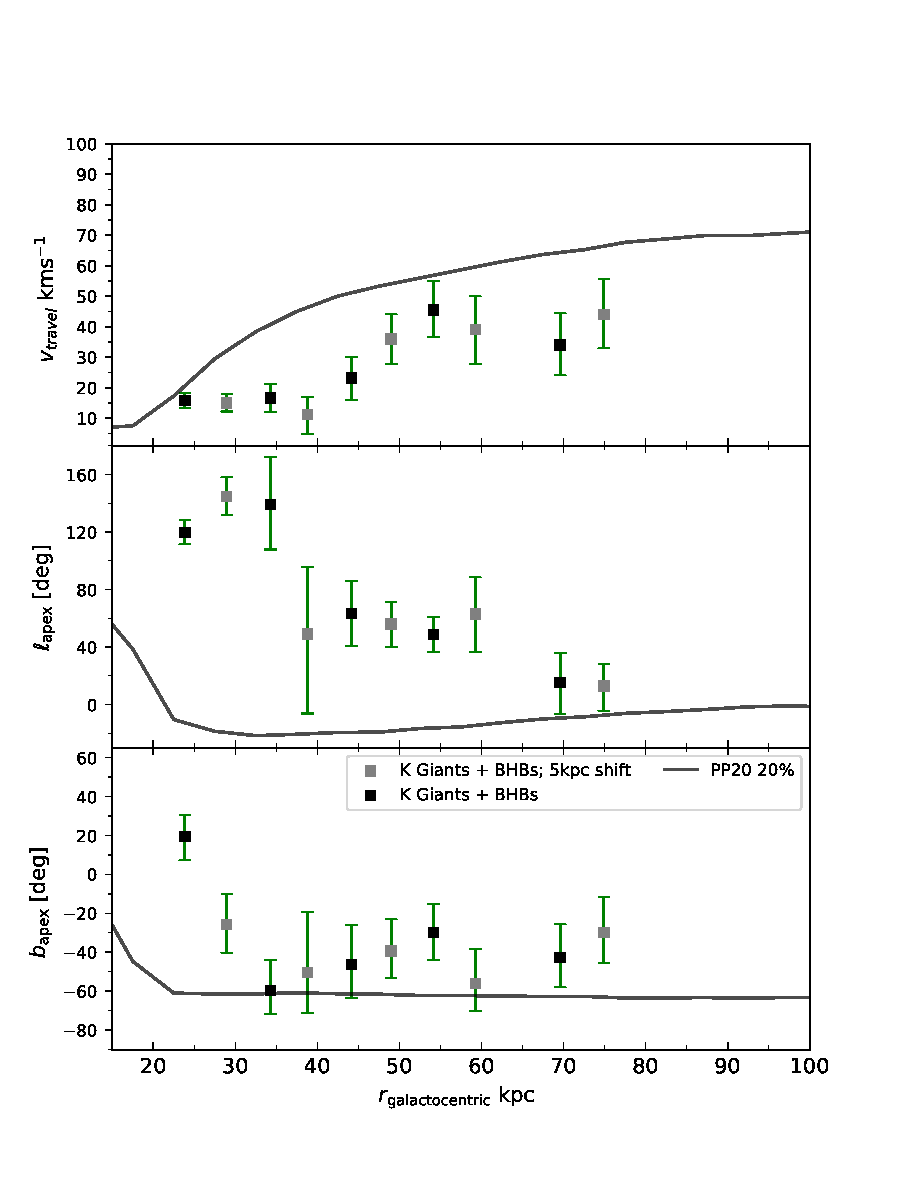}
    \caption{Tests of the effect of bin choices on the dipole parameters for the combined K giants and BHBs sample. This figure shows measured values of the dipole parameters when varying the bin centre by 5 kpc. Top panel: fitted travel velocity in each bin with uncertainties given as the standard deviation of the posterior chains of the parameters. Middle panel: Measured $\lapex$ values for stars in each bin, note that we restrict the apex longitude to be between $180^{\circ}$ and $-20^{\circ}$. Bottom panel: Measured $\bapex$ values for stars in each bin, where the range of latitude angles are limited to $-90^{\circ}$ and $60^{\circ}$ In each panel, we plot the measurement from the \citetalias{2020MNRASReflexPP20} simulation as a visual guide. The colours of the model curves correspond to different bin centres, where the grey points are the results of the shifted bin centre, and the black points are the original bins between 20 and 60 kpc. Error bars indicate the 1$\sigma$ width of the posterior distribution for each parameter.}
    \label{fig:D1-binvariation}
\end{figure}

\begin{figure}
    \includegraphics[width=\columnwidth]{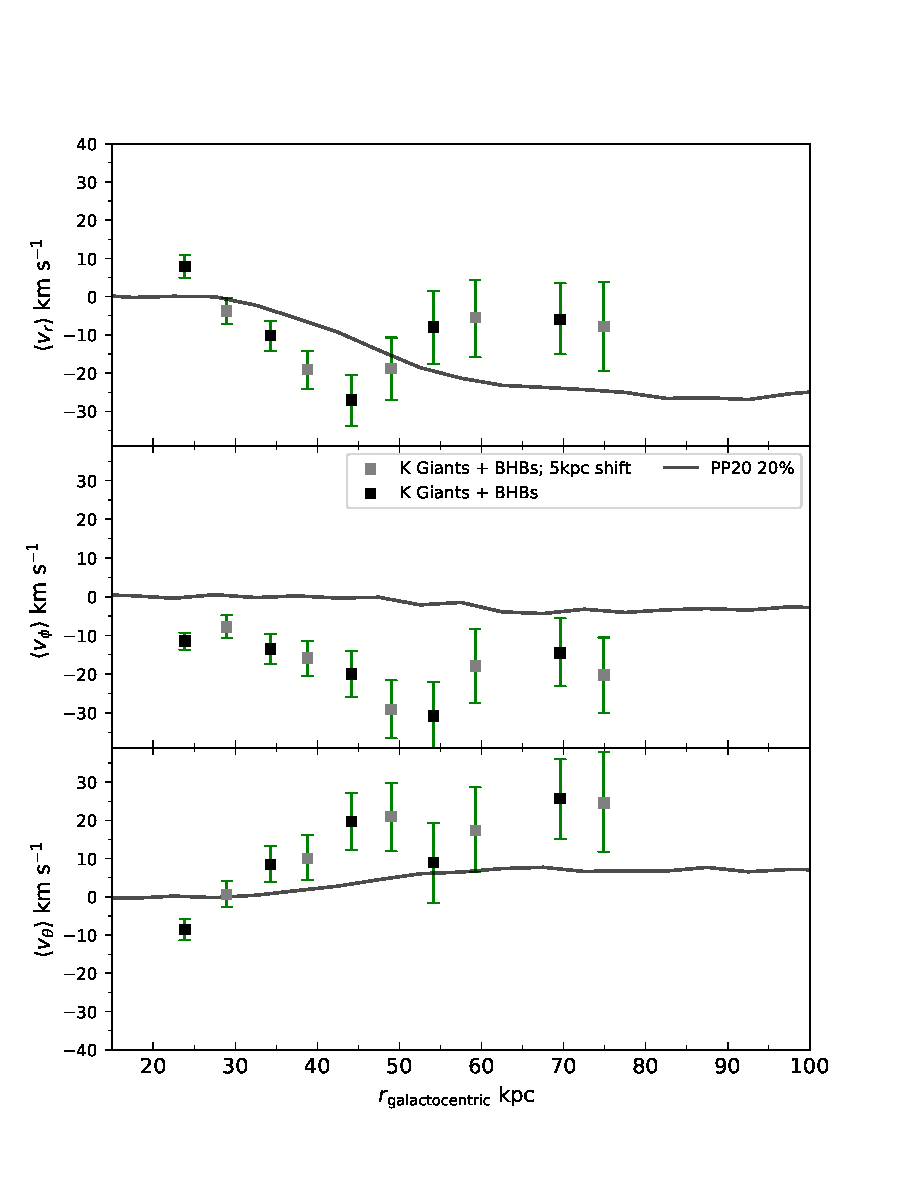}
    \caption{Tests of the effect of bin choices on the bulk motion parameters for the combined K giants and BHBs sample. Top panel: Mean halo motion in the radial direction.This figure shows measured values of the bulk motion parameters when varying the bin centre by 5 kpc. Middle panel: Mean halo motion in the azimuthal direction(cylindrical rotation). Bottom Panel: mean halo motion in the polar direction. In each panel, we plot the measurement from the \citetalias{2020MNRASReflexPP20} simulation as a visual guide. The colours of the model curves correspond to different bin centres, where the grey points are the results of the shifted bin centre, and the black points are the original bins between 20 and 60 kpc. Error bars indicate the 1$\sigma$ width of the posterior distribution for each parameter.}
    \label{fig:D2-binvariation}
\end{figure}

In this Appendix we vary the bin centres of the sample to check whether there are significant biases arising from the choice of bin centre. Figures~\ref{fig:D1-binvariation} and \ref{fig:D2-binvariation} show the results of tests with two different bin centres. We do not see evidence for significant changes when using different bins. Further, the re-centring of the bins exemplifies the continuous behaviour of the model parameters with distance.

As in Appendix~\ref{appendix:b}, to guide the eye between different figures, we include curves from the 20\% LMC model from \citetalias{2020MNRASReflexPP20}. While we show the tracks in Figures~\ref{fig:D1-binvariation} and \ref{fig:D2-binvariation}, the curves should not be interpreted as targets: the key result in this Appendix is again the difference between the different fit points at a given radius.
    
\bsp	
\label{lastpage}
\end{document}